\documentclass[12pt]{article}
\usepackage{setspace}
\usepackage{epsf}
\usepackage{natbib}
\usepackage{epsfig}
\usepackage{amsthm}
\usepackage{color}
\usepackage{amsmath}
\usepackage{amssymb}

\newcommand{\bm}[1]{\mbox{\boldmath $#1$}}
\newcommand{\mb}[1]{\mathbf{#1}}

\begin{document}

\doublespacing

\title{
  Bayesian Treed Gaussian Process Models with an Application to
  Computer Modeling
}
\author{
\vspace{-0.4cm}
Robert B. Gramacy and 
Herbert K. H. Lee\thanks{\noindent Robert Gramacy is Lecturer,
  Statistical Laboratory, University of Cambridge, UK (Email:
  bobby@statslab.cam.ac.uk) and Herbert Lee is Associate Professor,
  Department of Applied Mathematics and Statistics, University of
  California, Santa Cruz, CA 95064 (Email: herbie@ams.ucsc.edu).  
  The authors would like to thank William Macready for originating the
  collaboration with NASA and for his help with the project, 
  Thomas Pulliam and Edward Tejnil for their help with the NASA data,
  Tamara Broderick for her careful reading and detailed comments,
  and the editor, associate editor, and two anonymous
  referees for their helpful comments and suggestions.  This work was 
 partially supported by NASA awards 08008-002-011-000 and SC 2003028
 NAS2-03144, Sandia National Labs grant 496420, and National Science
 Foundation grants DMS 0233710 and 0504851.}\\
\date{}
}
\maketitle

\begin{abstract}
\noindent
Motivated by a computer experiment for the design of a rocket booster,
this paper explores nonstationary
modeling methodologies that couple stationary Gaussian processes with
treed partitioning.  Partitioning is a
simple but effective method for dealing with nonstationarity.  
The methodological developments and statistical
computing details which make this approach efficient are described in
detail.  In addition to providing an analysis of the rocket booster
simulator, our approach is demonstrated to be effective in 
other arenas.

\noindent
{\bf Key words:} computer simulator, recursive partitioning, nonstationary 
spatial model, nonparametric regression, heteroscedasticity
\end{abstract}


\newpage
\section{Introduction}
\label{sec:intro}
\vspace{-0.2cm}

Much of modern engineering design is now done through computer
modeling, which is both faster and more cost-effective than building
small-scale models, particularly in the earlier stages of design when
more scope for changes is desired.  As design proceeds, increasingly
sophisticated simulators may be created.  Our work here was motivated
by a simulator of a proposed rocket booster.  NASA relies heavily on
simulators for design, as wind tunnel experiments are quite expensive
and still not fully realistic of the range of flight experiences.  In
particular, one of the highly critical points in time for a rocket
booster is the moment that it re-enters the atmosphere.  Such
conditions are difficult to recreate in a wind tunnel, and it is
obviously impossible to run a standard physical experiment.  Thus to
learn about the behavior of the proposed rocket booster, NASA uses
computer simulation.

Simulators can involve large amounts of physical modeling, requiring
the numerical solution of complex systems of differential equations.
The NASA simulator was no exception, typically requiring between five
and twenty hours for a single run.  Thus, NASA was interested in
creating a statistical model of the simulator itself, an {\it
  emulator} or {\em surrogate model}, in the terminology of computer
modeling.  The standard approach in the literature for emulation is to
model the simulator output with a stationary smooth Gaussian process (GP)
\citep{sack:welc:mitc:wynn:1989,kennedy:ohagan:2001,sant:will:notz:2003}.
However, this approach proved to be inadequate for the NASA data.  In
particular, we were faced with problems of nonstationarity,
heteroscedasticity, and the size of the dataset.  Thus we introduce
here an expansion of GPs, based on the idea of Bayesian
partition models \citep{chip:geor:mccu:2002,deni:adam:holm:hand:2002},
which is able to address these issues.

GPs are conceptually straightforward, can easily accommodate prior
knowledge in the form of covariance functions, and can return
estimates of predictive confidence, which were desired by NASA.
However, we highlight three disadvantages of the standard form of a GP
which we had to confront on this dataset, and expect to encounter on a
wide range of other applications.  First, inference on the GP scales
poorly with the number of data points, $N$, typically requiring
computing time in $O(N^3)$ for calculating inverses of $N\times N$
covariance matrices.  Second, GP models are usually stationary in that
the same covariance structure is used throughout the entire input
space, which may be too strong of a modeling assumption.  Third, the
estimated predictive error (as opposed to the predictive mean value)
of a stationary model does not directly depend on the locally observed
response values.  Rather, the predictive error at a point depends only
on the locations of the nearby observations and on a global measure of
error that uses all of the discrepancies between observations and
predictions without regard for their distance from the point of
interest (because of the stationarity assumption).
(Section~\ref{sec:tgp:pred} provides more details, in particular note
that Eq.~(\ref{eq:predvar}) does not depend on ${\bf z}$.)  In many
real-world spatial and stochastic problems, such a uniform modeling of
uncertainty will not be desirable.  Instead, some regions of the space
will tend to exhibit larger variability than others.  On the other
hand, fully nonstationary Bayesian GP models
\citep[e.g.,][]{higd:swal:kern:1999,schmidt:2003} can be difficult to
fit, and are not computationally tractable for more than a relatively
small number of datapoints.  Further discussion of nonstationary
models is deferred until the end of Section~\ref{sec:intro:part}.

All of these shortcomings can be addressed by partitioning the input
space into regions, and fitting separate stationary GP models within
each region \citep[e.g.,][]{kim:mall:holm:2005}.  Partitioning
provides a relatively straightforward mechanism for creating a
nonstationary model, and can ameliorate some of the computational
demands by fitting models to less data.  A Bayesian model averaging
approach allows for the explicit estimation of predictive uncertainty,
which can now vary beyond the constraints of a stationary model.
Finally, an {\tt R} package with implementations of all of the models
discussed in this paper is available at\\
\verb!http://www.cran.r-project.org/web/packages/tgp/index.html!.
We note that by partitioning, we do not have any theoretical guarantee
of continuity in the fitted function.  However, as we demonstrate in
several examples, Bayesian model averaging yields mean fitted
functions that are typically quite smooth in practice, giving fits
that are indistinguishable from continuous functions except when the
data call for the contrary.  Indeed the ability to accurately model
possible discontinuities is a side benefit of this approach.

The rest of the paper is organized as follows.  Section~\ref{sec:lgbb}
describes the motivating data in further detail.
Section~\ref{sec:intro:relwork} provides some background material.
Section~\ref{sec:tgp} combines stationary GPs and treed partitioning
to create treed GPs, implementing a tractable nonstationary model for
nonparametric regression.  In Section~\ref{sec:lgbb:results} we return
to the analysis of the rocket booster data, and in
Section~\ref{sec:conclude} we conclude with some discussion.

\vspace{-0.2cm}
\section{The Langley Glide-Back Booster Simulation}
\label{sec:lgbb}
\vspace{-0.1cm}

The Langley Glide-Back Booster (LGBB) is a proposed rocket booster
under design at NASA.  Standard rocket boosters are created to be
reusable, assisting in the launch process and then parachuting back to
the Earth after their fuel has been exhausted.  Their return path is
planned so that they fall into the ocean, where they can be recovered
and reused.  The LGBB represents a new direction in booster design, as
it would have wings and a tail, looking somewhat similar to the space
shuttle.  The idea is that it would gracefully glide back down, rather
than plummeting into the ocean.  

The development of the booster is being done primarily through the use
of computer simulators.  The particular model \citep{rog03} with which
we were involved is based on computational fluid dynamics simulators
that numerically solve the relevant inviscid Euler equations over a
mesh of 1.4 million cells.  Each run of the Euler solver for a given
set of parameters can take 5-20 hours on the NASA computers.  The
simulator is theoretically deterministic, but the solver is typically
started with random initial conditions and does not always numerically
converge.  There is an automated check for convergence which is mostly
accurate, but some runs are marked as accepted despite their false
convergence, or they converge to a clearly inferior local mode.  For those
runs that fail the automated convergence check, the solver is
restarted at a different set of randomly chosen initial conditions.
Our NASA collaborators have commented that input configurations
arbitrarily close to one another can fail to achieve the same
estimated convergence, even after satisfying the same stopping
criterion.  Thus neither simple interpolation of the data nor a
Gaussian process model without an error term will be adequate, as
smoothing will be necessary to reduce the impact of the inaccurate
runs.

The simulator models the forces felt by the vehicle at the moment it
is re-entering the atmosphere.  As a free body in space, there are six
degrees of freedom, so the six relevant forces are lift, drag, pitch,
side-force, yaw, and roll.  For this project, the interest focused
just on the lift force, as that is the most important one for keeping
a vehicle aloft.  The inputs to the simulator are the speed of the
vehicle at re-entry (measured by Mach number), the angle of attack
(the alpha angle), and the sideslip angle (the beta angle).  Thus the
primary goal is to model the lift force as a function of speed, alpha,
and beta.  The sideslip angle is quantized in the experiments, so it
is run only at six particular levels.  Speed ranges from Mach 0 to 6,
and the angle of attack, alpha, varies from negative five to thirty
degrees.  The simulator was run at 3041 locations, over a combination
of three hand-designed grids.  The first grid was relatively coarse
and was equally spaced over the whole region of interest.  Two
successively finer grids on smaller regions primarily around Mach one
were further run, as the initial run showed that the most interesting
part of the input space was generally around the sound barrier.  This
makes sense because the physics in the simulator comes from two
completely different regimes, a subsonic regime for speeds less than
Mach one, and a supersonic regime for speeds greater than Mach one.
What happens close to and along the boundary is the most difficult
part of the simulation.

\begin{figure}[ht!]
\begin{center}
\includegraphics[angle=-90,scale=0.33,trim=20 185 50 175]{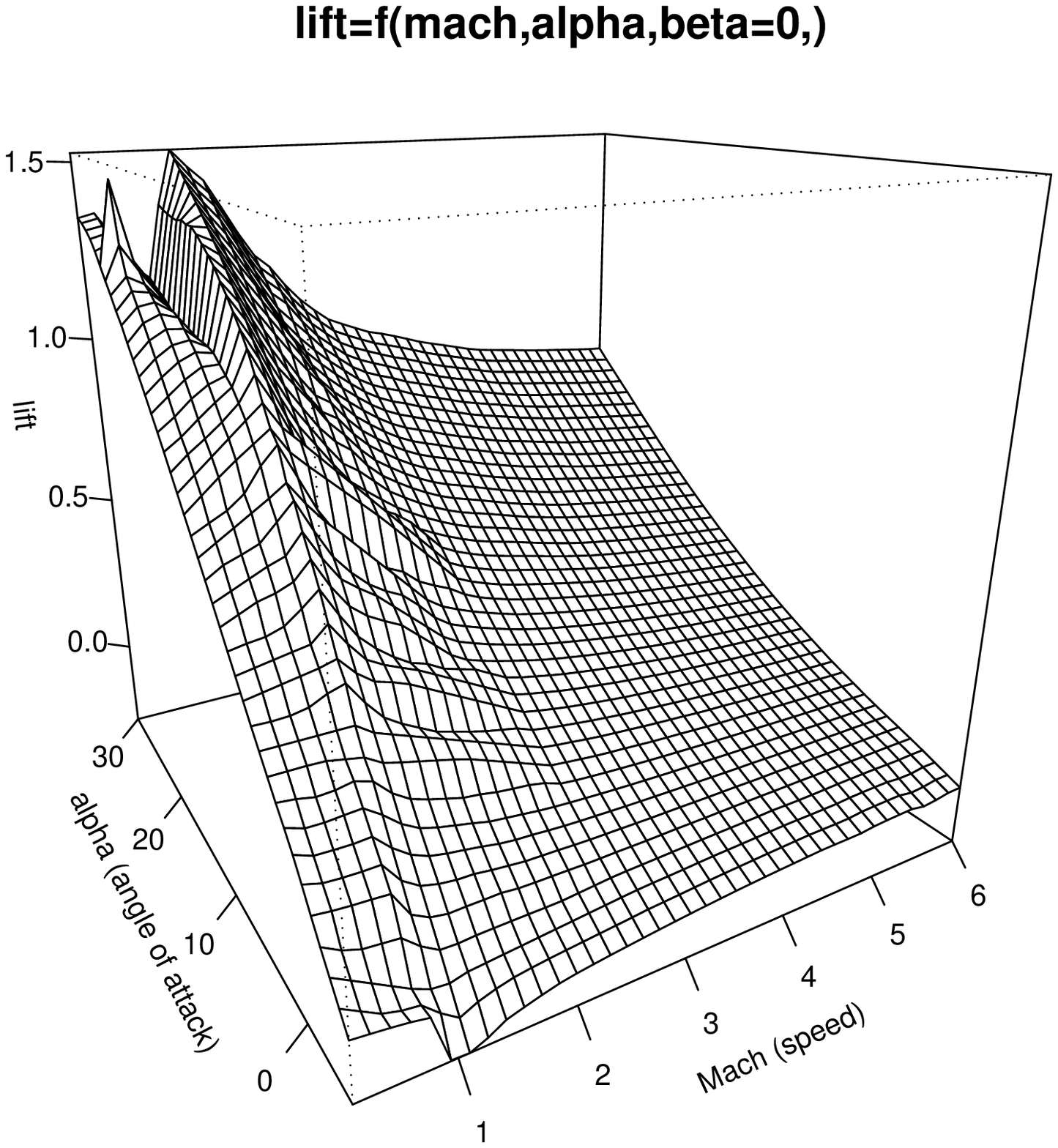} 
\includegraphics[angle=-90,scale=0.33,trim=20 185 50 175]{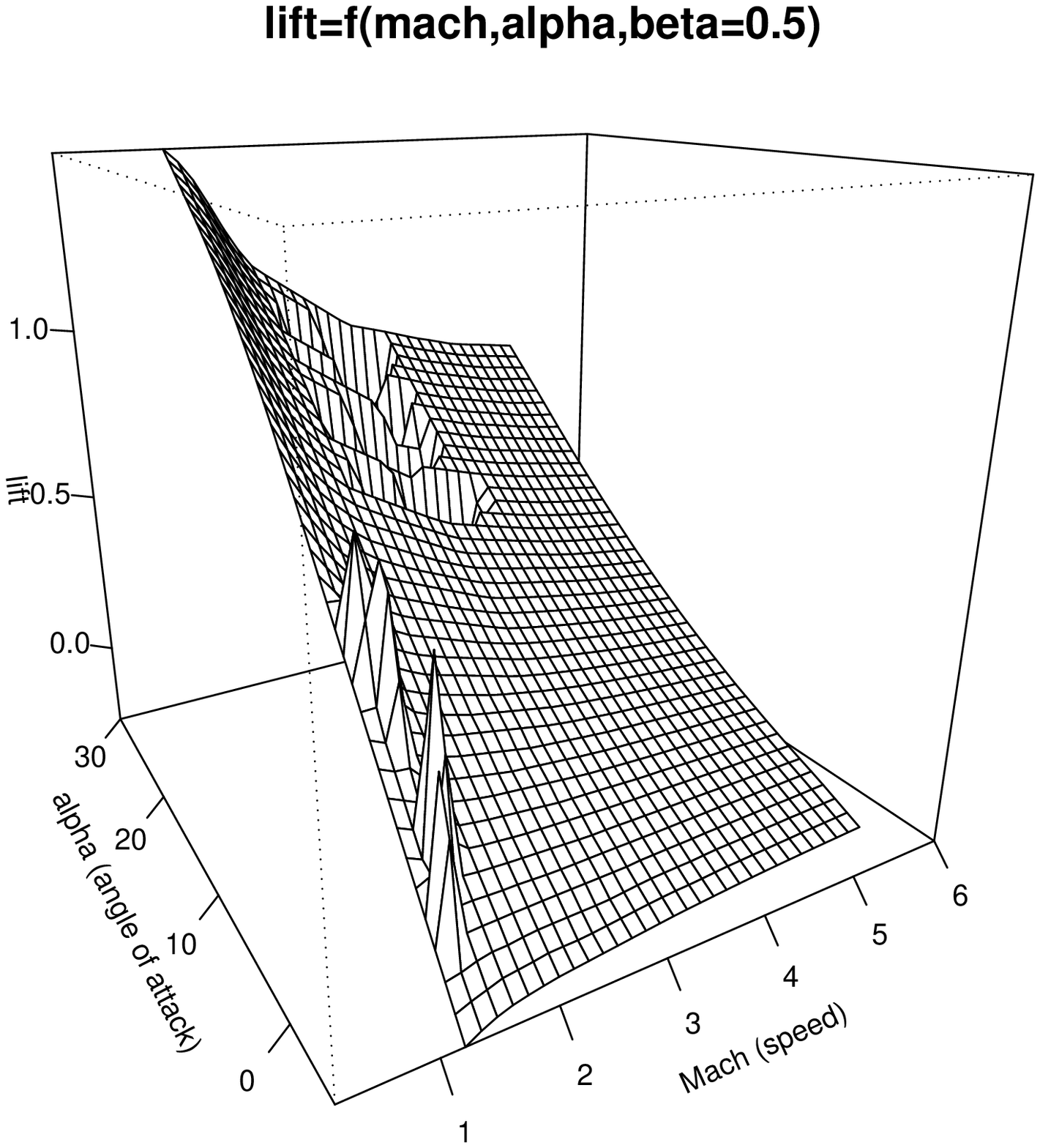}
\includegraphics[angle=-90,scale=0.33,trim=20 185 50 185]{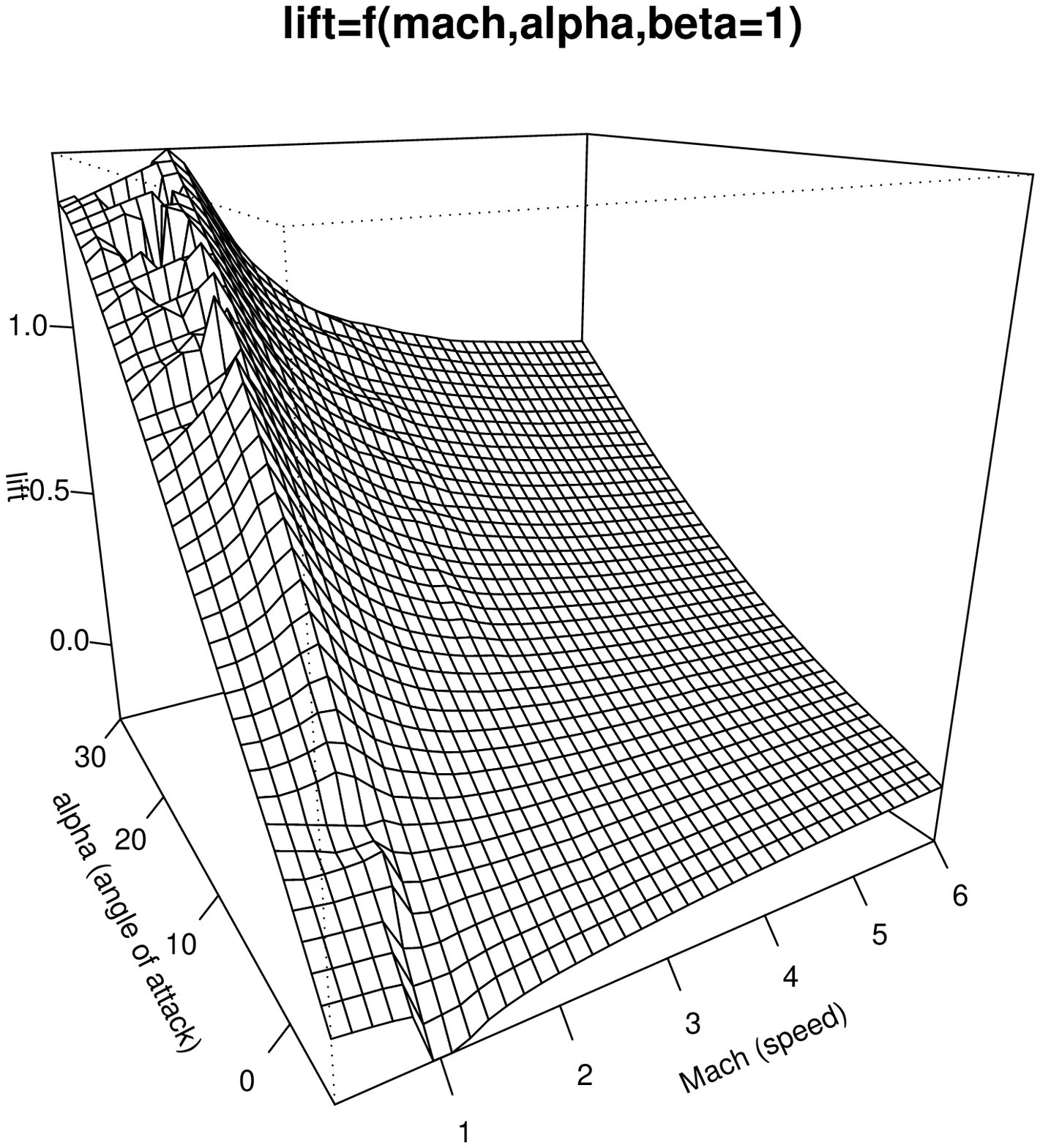}
\includegraphics[angle=-90,scale=0.33,trim=20 185 75 175]{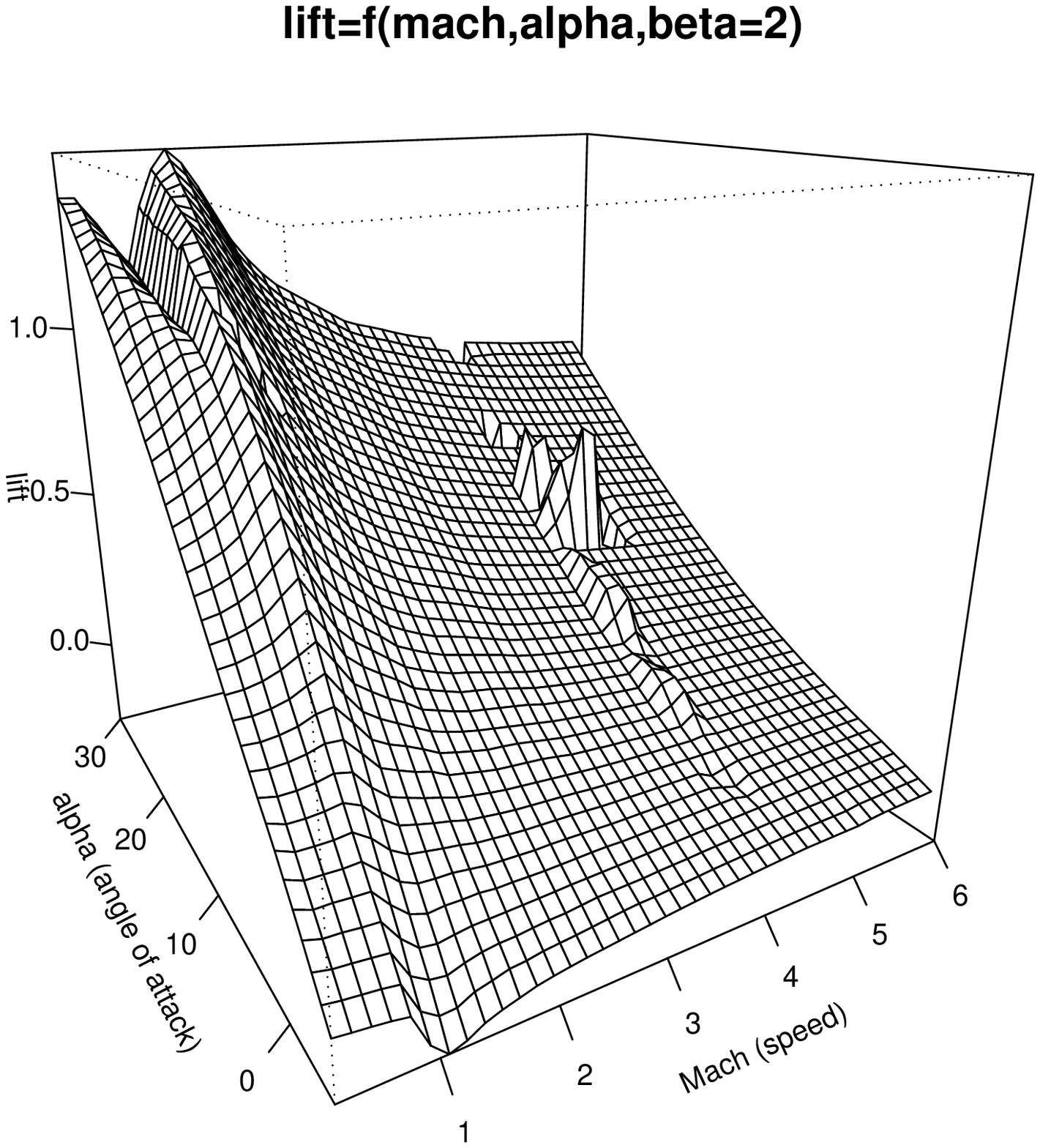}
\includegraphics[angle=-90,scale=0.33,trim=20 185 75 175]{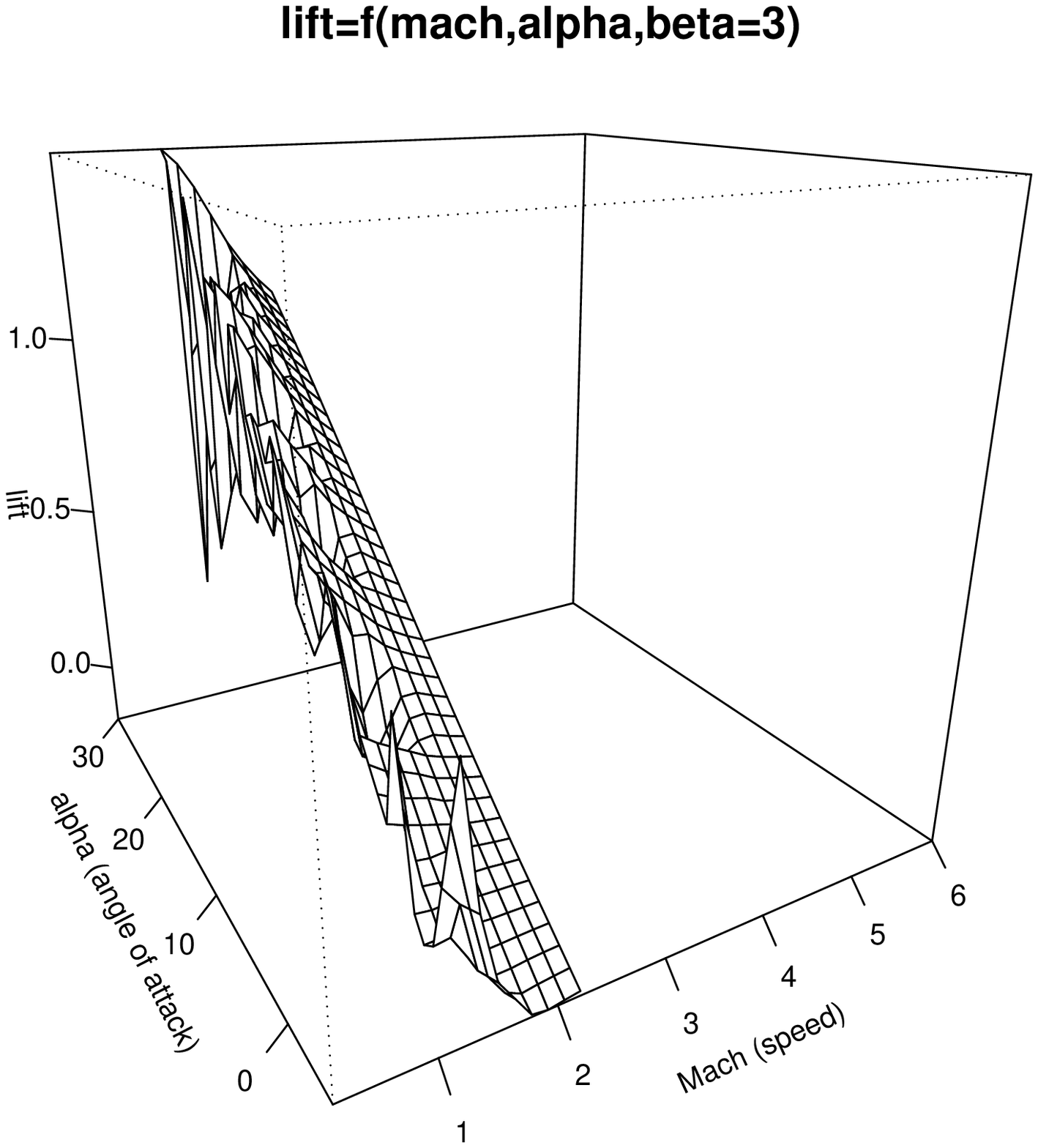}
\includegraphics[angle=-90,scale=0.33,trim=20 185 75 185]{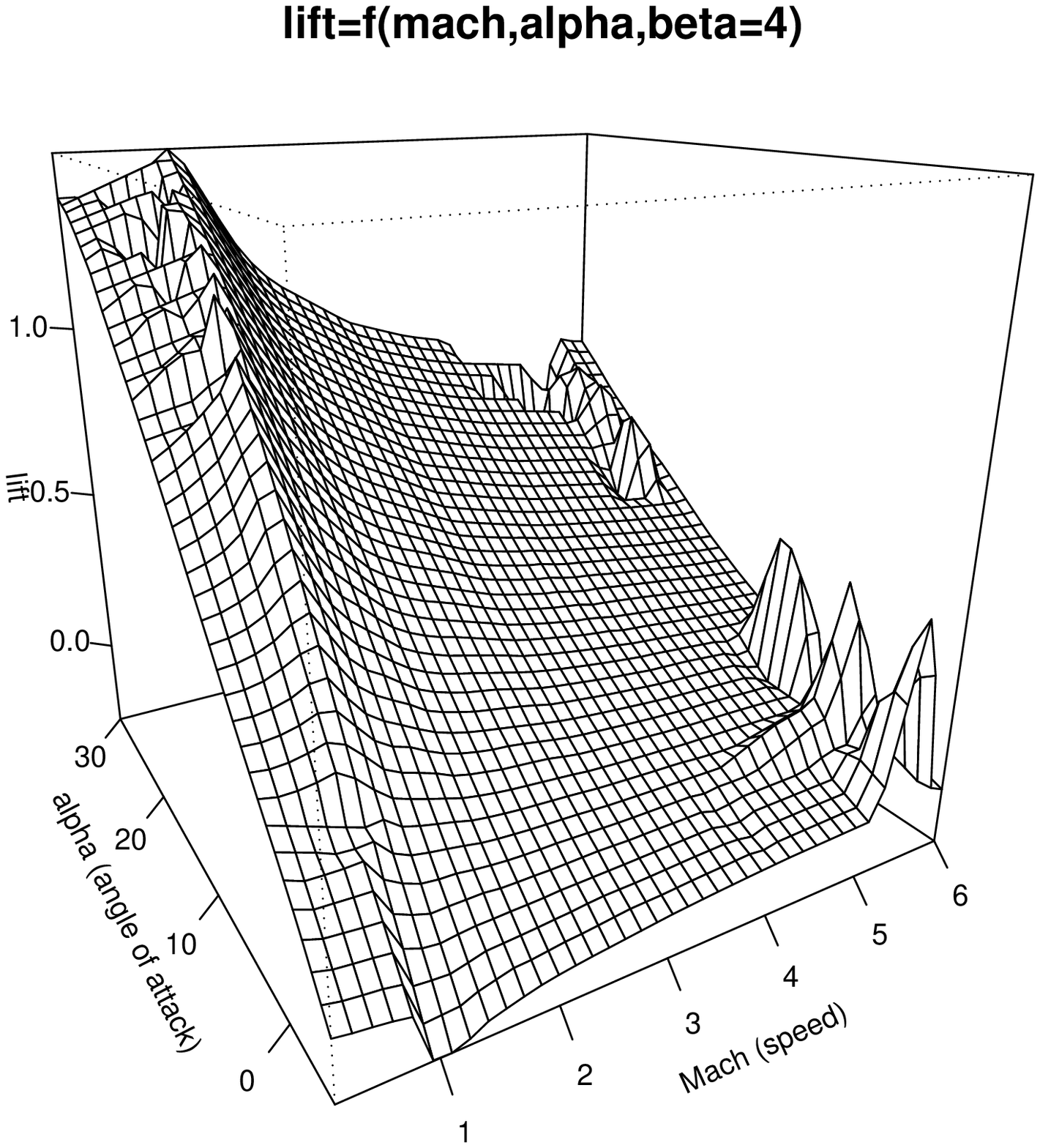} 
\end{center}
\vspace{-0.2cm}
\caption{Interpolation of lift by speed and angle of attack
  for all sideslip levels.
  Note that for levels 0.5 and 3 (center), Mach ranges only in $(1,5)$
  and $(1.2, 2.2)$.}
\label{f:data0}
\vspace{-0.2cm}
\end{figure}

The upper-left plot in Figure~\ref{f:data0} shows an interpolation of the
simulator output for the lift surface as a function of speed and angle
of attack, when the sideslip angle is zero.  The primary feature of
this plot is the large ridge which appears at Mach one and larger
angles of attack.  The transition from subsonic to supersonic is a
sharp one, and it is not clear whether one would want to use a
continuous model or to introduce a discontinuity.  While much of the
surface is quite smooth, parts of the surface, particularly around
Mach one, are less smooth.  So it is clear that the standard computer
modeling assumption of a stationary process will not work well here.
We will need a method that allows for a nonstationary formulation, yet that can
still produce uncertainty estimates, and is computationally feasible
to fit on a dataset of this size.

One other feature of the data that appears in Figure~\ref{f:data0} is
the issue of numerical convergence.  In the upper-left corner of the
upper-left plot (high angle of attack, low speed), there is a single
spike that looks out of place.  Our collaborators at NASA believe this
to be a result of a false convergence by the simulator, so we would
want our surrogate model to smooth this one point out.  This stands in
contrast to most computer modeling problems, where one wants to
interpolate the deterministic simulator without smoothing.  Here we
require smoothing to compensate for problems with the simulator.

The other plots of Figure~\ref{f:data0} show the issue of false
convergence more strongly for other sideslip settings.  In the center
plots (levels one-half and three), there are noisy depressions in the
surface for moderate speed and high angle of attack.  Because this
feature is not seen in the other plots by sideslip angle, one may
suspect that this region could be showing more numerical instability
than signal.  Thus, there is a need to combine information across the
levels in order to smooth out numerical problems with the simulator.
Note that because no subsonic inputs were sampled for these slices,
the ridge around Mach one does not appear in these two plots.

For sideslip levels of one, two, and four the surface again appears to
be most interesting around Mach one.  But instead of a clean ridge at
levels one and four it is noisy, especially at high angles of attack.
It is not clear if this variability is due to false convergence of the
simulator, inadequate coverage in the design, or if the boundary
really is this complicated.  The NASA scientists have postulated that
the instabilities are more likely to be numerical, rather than
structural, but we will want our surrogate model to capture this
uncertainty.

Also of concern are the deviations from the smooth trend at high
speeds (particularly for level four), with upward deviations at low
angles of attack and downward deviations at high angles.  Again, it is
suspected that these are the result of false numerical convergence of
the simulator, but we cannot rule out a priori that the physical
system itself becomes unstable at higher speeds.
So we desire smoothing, but with an appropriate local estimate of
uncertainty.  Fitting with a single stationary GP would
give uncertainty estimates that were fairly uniform across the space,
because of the assumption of stationarity.  Thus we turn to a
partitioning approach.

Understanding the mean surface is important for the engineers for
several reasons.  First, they may discover potential problems with the
design, which could lead to structural changes in the design.  Second,
they will need to determine the optimal flight conditions so they can
plan the flight trajectories.  Third, they need to be able to make
contingency plans in case problems arise during a mission.  If some of
the stabilizing rockets fail and the vehicle must re-enter at an
unplanned angle or speed, they will need to be able to map out its new
trajectory and adjust the process as necessary.  The engineers are
interested in not just the mean surface, but also the uncertainty
associated with prediction, because these uncertainties are not
constant across the surface.

\vspace{-0.2cm}
\section{Related work}
\label{sec:intro:relwork}
\vspace{-0.1cm}

Our approach to nonparametric and semiparametric nonstationary
modeling combines standard GPs and treed partitioning within the
context of Bayesian hierarchical modeling and model averaging.  We
assume that the reader is familiar with the basic concepts of Bayesian
inference via Markov chain Monte Carlo \citep[e.g.,][]{gilks:1996}.
An introduction to GPs and treed partition modeling follows.

\vspace{-0.1cm}
\subsection{Stationary Gaussian Processes}
\label{sec:intro:gp}
\vspace{-0.1cm}

A common specification of stochastic processes for spatial data, of
which the stationary Gaussian process (GP) is a particular case,
specifies that model outputs (responses) $z$, depend on multivariate
inputs (explanatory variables) $\mb{x}$, as $z(\mb{x}) =
\bm{\beta}^\top \mb{f}(\mb{x}) + w(\mb{x})$ where $\bm{\beta}$ are
linear trend coefficients, $w(\mb{x})$ is a zero mean random process
with covariance $C(\mb{x}, \mb{x}') = \sigma^2 K(\mb{x}, \mb{x}')$,
$\mb{K}$ is a correlation matrix, and $\bm{\beta}$ is independent of
$w$ in the prior.  Low-order polynomials are
sometimes used instead of the simple linear mean $\bm{\beta}^\top
\mb{f}(\mb{x})$, or the mean process is specified generically, often
as $\xi(\mb{x}, \bm{\beta})$ or $\xi(\mb{x})$.

Formally, a Gaussian process is a collection of random variables
$Z(\mb{x})$ indexed by $\mb{x}$, having a jointly Gaussian
distribution for any finite subset of indices~\citep{stein:1999}.  It
is specified by a mean function $\mu(\mb{x}) =
E\bigl(Z(\mb{x})\bigr)$ and a correlation function
$K(\mb{x},\mb{x}') = \frac{1}{\sigma^2} E\bigl( [Z(\mb{x}) -
\mu(\mb{x})] [Z(\mb{x}') - \mu(\mb{x'})]^\top \bigr)$.
We assume that the correlation function can be written in the form
\vspace{-0.2cm}
\begin{equation} 
K(\mb{x}_j, \mb{x}_k|g) =
        K^*(\mb{x}_j, \mb{x}_k) + {g} \delta_{j,k}. \label{eq:cor}
\vspace{-0.2cm}
\end{equation} 
where $\delta_{\cdot,\cdot}$ is the Kronecker delta function and
$K^*$ is a proper underlying parametric correlation function.
The $g$ term in Eq.~(\ref{eq:cor}) is referred to as the nugget.  It must
always be positive $(g>0)$, and it serves two 
purposes.  First, it provides a mechanism for introducing
measurement error into the stochastic process.  It arises when
considering a model of the form
$
Z(\mb{x}) = \xi(\mb{x}, \bm{\beta}) + w(\mb{x}) + \eta(\mb{x}), 
$ 
where $w(\cdot)$ is a process with correlations
governed by $K^*$, and $\eta(\cdot)$ is simply Gaussian
noise.  Second, the
nugget helps prevent $\mb{K}$ from becoming numerically
singular.  Notational convenience and conceptual congruence motivates
referral to $\mb{K}$ as a correlation matrix, even though the nugget
term ($g$) forces $K(\mb{x}_i,\mb{x}_i)>1$.  There is an isomorphic
model specification wherein $\mb{K}$ depicts proper correlations.
Under both specifications $K^*$ does indeed define a valid correlation
matrix $\mb{K}^*$.

The correlation functions $K^*(\cdot, \cdot)$ are typically
specified through a low dimensional parametric structure, which also
guarantees that they are symmetric and positive semi-definite.
Here we focus on the power
family, although our methods are clearly extensible to other families,
such as the Mat\'{e}rn class \citep{mate:1986}.  Further discussion of
correlation structures can be found in 
\citet{abraham:1997} or \citet{stein:1999}.  The power family of correlation
functions includes the simple isotropic parameterization
\vspace{-0.3cm}
\begin{equation} 
        K^*(\mb{x}_j, \mb{x}_k|d)  =
        \exp\left\{-\frac{||\mb{x}_j - \mb{x}_k||^{p_0}}{d} \right\}, 
        \label{eq:pow} 
\end{equation} 
where $d>0$ is a single range parameter and $p_0 \in (0,2]$ determines
the smoothness of the process.  Thus the correlation of two points
depends only on the Euclidean distance $||\mb{x}_j - \mb{x}_k||$
between them.  A straightforward enhancement to the isotropic power
family is to employ a separate range parameter $d_i$ in each dimension
($i=1,\dots,m_X$).  The resulting correlation function is still
stationary, but no longer isotropic:
\vspace{-0.2cm}
\begin{equation} 
        K^*(\mb{x}_j, \mb{x}_k|\mb{d}) = 
        \label{e:cor_d} \exp\left\{ - \sum_{i=1}^{m_X}
        \frac{|x_{ij} - x_{ik}|^{p_0}}{d_{i}}\right\} \,.
\end{equation} 


\vspace{-0.2cm}
\subsection{Treed Partitioning}
\label{sec:intro:part}
\vspace{-0.1cm}

Many spatial modeling problems require more flexibility than is
offered by a stationary GP.  One way to achieve a more flexible,
nonstationary, process is to use a partition model---a meta-model
which divides up the input space and fits different base models to
data independently in each region.  Treed partitioning is one possible
approach.

Treed partition models typically divide up the input space by making
binary splits on the value of a single variable (e.g., $x_1 > 0.8$) so
that partition boundaries are parallel to coordinate axes.
Partitioning is recursive, so each new partition is a sub-partition of
a previous one.  For example, a first partition may divide the space
in half by whether the first variable is above or below its midpoint.
The second partition will then divide only the space below (or above)
the midpoint of the first variable, so that there are now three
partitions (not four).  Since variables may be revisited, there is no
loss of generality by using binary splits, as multiple splits on the
same variable will be equivalent to a non-binary split.  In each
partition (leaf of the tree), an independent model is
applied.  Classification and Regression Trees (CART)~\citep{brei:1984}
are an example of a treed partition model.  CART, which fits a 
constant surface in each leaf, has become popular because of its ease of use,
clear interpretation, and ability to provide a good fit in many cases.

The Bayesian approach is straightforward to apply to
CART~\citep{chip:geor:mccu:1998,deni:mall:smit:1998}, provided that
one can specify a meaningful prior for the size of the tree.  We
follow Chipman et al.~(1998)~\nocite{chip:geor:mccu:1998} who specify
the prior through a tree-generating process and enforce a minimum
amount of data in order to infer the parameters in each partition.
Starting with a null tree (all data in a single region), a leaf node
$\eta \in \mathcal{T}$, representing a region of the input space,
splits with probability $a (1 + q_\eta)^{-b}$, where $q_\eta$ is the
depth of $\eta\in\mathcal{T}$ and $a$ and $b$ are parameters chosen to
give an appropriate size and spread to the distribution of trees.
Further details are available in the Chipman et al.~papers.  
For our models, we have found that default values of $a=0.5$ and $b=2$
often work well in practice, although in any particular problem prior
knowledge may call for other values.
The prior for the splitting process involves first choosing the
splitting dimension $u$ from a discrete uniform, and then the split
location $s$ is chosen uniformly from a subset of the locations
$\mb{X}$ in the $u^{\mbox{\tiny th}}$ dimension.  Integrating out
dependence on the tree structure $\mathcal{T}$ can be accomplished via
Reversible-Jump (RJ) MCMC as further described in
Section~\ref{sec:tgp:tree}.

Chipman et al.~(2002)~\nocite{chip:geor:mccu:2002} generalize Bayesian
CART to create the Bayesian treed linear model (LM) by fitting
hierarchical LMs at the leaves of the tree.  In Section~\ref{sec:tgp}
we generalize further by proposing to fit stationary GPs in each of
the leaves of the tree.  This approach bears some similarity to that
of \citet{kim:mall:holm:2005}, who fit separate GPs in each element of
a Voronoi tessellation.  The treed GP approach is better geared toward
problems with a smaller number of distinct partitions, leading to a
simpler overall model.  Voronoi tessellations allow an intricate
partitioning of the space, but have the trade-off of added complexity
and can produce a final model that is difficult to interpret.  The
tessellation approach also has the benefit of not being restricted to
axis-aligned partitions (although in some cases, simple
transformations such as rotating the data will suffice to allow
axis-aligned partitions).  A nice review of Bayesian partition
modeling is provided by \citet{deni:adam:holm:hand:2002}.


Other approaches to nonstationary modeling include those which use
spatial deformations and process convolutions.  The idea behind the
spatial deformation approach is to map nonstationary inputs in the
original, geographical, space into a dispersion space wherein the
process is stationary.  \citet{samp:gutt:1992} use thin-plate spline
models and multidimensional scaling (MDS) to construct the mapping.
\citet{dam:samp:gutt:2001} explore a similar methodology from a
Bayesian perspective.  \citet{schmidt:2003} also take the Bayesian
approach, but put a Gaussian process prior on the mapping.  The
process convolution approach
\citep{higd:swal:kern:1999,fuen:2002,Paci:2003} proceeds by
allowing the convolution kernels 
to vary smoothly in parameterization as an unknown function of their
spatial location.
A common theme among such nonstationary models is the introduction of
meta-structure which ratchets up the flexibility of the model,
ratcheting up the computational demands as well.  This is in stark
contrast to the treed approach that introduces a structural mechanism,
the tree $\mathcal{T}$, that actually reduces the computational burden
relative to the base model, e.g., a GP, because smaller correlation
matrices are inverted.  A key difference is that these alternative
approaches strictly enforce continuity of the process, which requires
much more effort than the treed approach.

\vspace{-0.2cm}
\section{Treed Gaussian process models}
\label{sec:tgp}
\vspace{-0.1cm}

Extending the partitioning ideas of Chipman et al.~(1998,
2002)\nocite{chip:geor:mccu:1998,chip:geor:mccu:2002} for simple
Bayesian treed models, we fit stationary GP models with linear trends
independently within each of $R$ regions, $\{r_\nu\}_{\nu=1}^R$,
depicted at the leaves of the tree $\mathcal{T}$, instead of constant
(1998) or linear (2002) models.  The tree is averaged out by
integrating over possible trees, using RJ-MCMC~\citep{rich:gree:1997},
with the tree prior specified through a tree-generating process.
Prediction is conditioned on the tree structure, and is averaged over
in the posterior to get a full accounting of uncertainty.

\vspace{-0.2cm}
\subsection{Hierarchical Model}
\label{sec:tgp:hier}
\vspace{-0.1cm}

A tree $\mathcal{T}$ recursively partitions the input space into into
$R$ non-overlapping regions: $\{r_\nu\}_{\nu=1}^R$.  Each region
$r_\nu$ contains data $D_\nu = \{\mb{X}_\nu, \mb{Z}_\nu\}$, consisting
of $n_\nu$ observations.  Let $m\equiv m_X+1$ be number of covariates
in the design (input) matrix $\mb{X}$ plus an intercept.  For each
region $r_\nu$, the hierarchical generative GP model is
\vspace{-0.2in}
\begin{align} 
\mb{Z}_\nu | \bm{\beta}_\nu, \sigma^2_\nu, \mb{K}_\nu &\sim 
N_{n_\nu}(\mb{\mb{F}}_\nu \bm{\beta}_\nu, \sigma^2_\nu \mb{K}_\nu), 
   \nonumber 
& \bm{\beta}_0 &\sim N_m(\bm{\mu}, \mb{B}) \\ 
\bm{\beta}_\nu | \sigma^2_\nu, \tau^2_\nu, \mb{W}, 
\bm{\beta}_0 &\sim N_m(\bm{\beta}_0,\sigma^2_\nu \tau^2_\nu \mb{W})
& \tau^2_\nu &\sim IG(\alpha_\tau/2, q_\tau/2), \label{eq:model} \\ 
\sigma^2_\nu &\sim IG(\alpha_\sigma/2, q_\sigma/2), \nonumber
& \mb{W}^{-1} &\sim W((\rho \mb{V})^{-1}, \rho), \nonumber
\vspace{-0.3cm}
\end{align} 
with $\mb{F}_\nu = (\mb{1}, \mb{X}_\nu)$, and $\mb{W}$ is an $m \times
m$ matrix.  The $N$, $IG$, and $W$ are the (Multivariate) Normal,
Inverse-Gamma, and Wishart distributions, respectively.  Hyperparameters
$\bm{\mu}, \mb{B},\mb{V},\rho, \alpha_\sigma, q_\sigma, \alpha_\tau,
q_\tau$ are treated as known.  The model (\ref{eq:model}) specifies a
multivariate normal likelihood with linear trend coefficients
$\bm{\beta}_\nu$, variance $\sigma^2_\nu$, and $n_\nu\times n_\nu$
correlation matrix $\mb{K}_\nu$.  The coefficients $\bm{\beta}_\nu$
are believed to have come from a common unknown mean $\bm{\beta}_0$
and region-specific variance $\sigma^2_\nu\tau^2_\nu$.  There is no
explicit mechanism in the model (\ref{eq:model}) to ensure that the
process near the boundary of two adjacent regions is continuous across
the partitions depicted by $\mathcal{T}$. However, the model can
capture smoothness through model averaging, as will be discussed in
Section~\ref{sec:tgp:pred}.  In our work with models for physical
processes, we frequently encounter problems with phase transitions
where the response surface is not smooth at the boundary between
distinct physical regimes, such as subsonic vs.~supersonic flight of
the rocket booster, so we view the ability to fit a discontinuous
surface as a feature of this model.

The GP correlation structure $K_\nu(\mb{x}_j, \mb{x}_k) =
K^*_\nu(\mb{x}_j, \mb{x}_k) + {g}_\nu \delta_{j,k}$ generating
$\mb{K}_\nu$ for each partition $r_\nu$ takes $K_\nu^*$ to be from the
isotropic power family (\ref{eq:pow}), or separable power family
(\ref{e:cor_d}), with a fixed power $p_0$, but unknown (random) range
and nugget parameters.  However, since most of the following
discussion holds for $K^*_\nu$ generated by other families, as well as
for unknown $p_0$, we shall refer to the correlation parameters
indirectly via the resulting correlation matrix $\mb{K}$, or function
$K(\cdot, \cdot)$.  For example, $p(\mb{K}_\nu)$ can represent either
of $p(d_\nu,g_\nu)$ or $p(\mb{d}_\nu,g_\nu)$, etc.  Priors that
encode a preference for a model with a nonstationary global covariance
structure are chosen for parameters to $K^*_\nu$ and $g_\nu$.  In particular, 
we propose a mixture of Gammas prior for $d$: \vspace{-0.2cm}
\begin{equation}
p(d,g) = p(d)\times p(g)
= p(g)\times \frac{1}{2}[G(d|\alpha=1,\beta=20) +
G(d|\alpha=10,\beta=10)].
\label{eq:dprior}
\end{equation}
It gives roughly equal mass to small $d$ representing a population of
GP parameterizations for wavy surfaces, and a separate population for
those which are quite smooth or approximately linear.   
We take the prior for $g$ to be Exp$(\lambda)$.  Alternatively, one
could encode the prior as $p(d,g) = p(d|g)p(g)$ and then use a
reference prior \citep{berg:deol:sans:2001} for $p(d|g)$. We
prefer the more deliberate mixture specification both because of its
modeling implications, as well as its ability to interface well with
limiting linear models \citep{gra:lee:2008b}.  

It may also be sensible to define the prior for $\{\mb{K}, \sigma^2,
\tau^2\}_\nu$ hierarchically, depending on parameters $\bm{\gamma}$
(not indexed by $\nu$), similar to how the population of
$\bm{\beta}_\nu$ parameters is given a common prior in terms of
$\bm{W}$ and $\bm{\beta}_0$ in (\ref{eq:model}).

\vspace{-0.3cm}
\subsection{Estimation}
\label{sec:tgp:est}
\vspace{-0.2cm}

The data $D_\nu = \{\mb{X},\mb{Z}\}_\nu$ are used to update the GP
parameters $\bm{\theta}_\nu \equiv \{\bm{\beta}, \sigma^2, \mb{K},
\tau^2\}_\nu$, for $\nu=1,\dots,R$.  Conditional on the tree $\mathcal{T}$,
the full set of parameters is denoted as $\bm{\theta} = \bm{\theta}_0
\cup \bigcup_{\nu=1}^R \bm{\theta}_\nu$, where
$\bm{\theta}_0=\{\mb{W}, \bm{\beta}_0, \bm{\gamma}\}$ denotes
upper-level parameters from the hierarchical prior that are also
updated.  Samples from the posterior 
distribution of $\bm{\theta}$ are gathered using Markov chain Monte
Carlo (MCMC) by first conditioning on the hierarchical prior parameters
$\bm{\theta}_0$ and drawing $\bm{\theta}_\nu | \bm{\theta}_0$ for
$\nu_1,\dots, \nu_R$, and then $\bm{\theta}_0$ is drawn as
$\bm{\theta}_0|\bigcup_{\nu=1}^R \bm{\theta}_\nu$.  Section
\ref{sec:tgp:gp} gives the details.  All parameters can be sampled
with Gibbs steps, except those that parameterize the covariance
function $K(\cdot, \cdot)$, e.g., $\{d, g\}_\nu$, which require
Metropolis-Hastings (MH) draws.  Section \ref{sec:tgp:tree} shows how
RJ-MCMC is used to gather samples from the joint posterior of
$(\bm{\theta}, \mathcal{T})$ by alternately drawing $\bm{\theta} |
\mathcal{T}$ and $\mathcal{T} | \bm{\theta}$.

\vspace{-0.2cm}
\subsubsection{GP parameters given a tree ($\mathcal{T}$)}
\label{sec:tgp:gp}
\vspace{-0.2cm}

Conditional conjugacy allows Gibbs sampling for most parameters.  
Full derivations of the following equations are available in
\citet{gramacy:2005}.  
The linear regression parameters $\bm{\beta}_\nu$ and prior mean
$\bm{\beta}_0$ both have multivariate normal
full conditionals: $\bm{\beta}_\nu | \mbox{rest} \sim
N_m(\tilde{\bm{\beta}}_\nu, \sigma_\nu^2 \mb{V}_{\tilde{\beta}_\nu})$,
and $\bm{\beta}_0 | \mbox{rest} \sim N_m(\tilde{\bm{\beta}}_0,
\bm{V}_{\tilde{\beta}_0})$, where
\vspace{-0.3cm}
\begin{align} 
\mb{V}_{\tilde{\beta}_\nu} &= \label{eq:betavar} 
(\mb{F}_\nu^\top\mb{K}_\nu^{-1}\mb{F}_\nu \!+\! \mb{W}^{-1}/\tau_\nu^2)^{-1}, &
  \tilde{\bm{\beta}}_\nu &= \mb{V}_{\tilde{\beta}_\nu}
  (\mb{F}_\nu^\top\mb{K}_\nu^{-1} \mb{Z}_\nu \!+\!
  \mb{W}^{-1}\bm{\beta}_0/\tau_\nu^2),\\ 
\mb{V}_{\tilde{\beta}_0} &= \left(\mb{B}^{-1}\! \!+\! \mb{W}^{-1}\!
   \textstyle \sum_{\nu=1}^R (\sigma_\nu\tau_\nu)^{-2}\right)^{-1}\!\!\!, &
\tilde{\bm{\beta}}_0 &=
\mb{V}_{\tilde{\beta}_0} \left(\mb{B}^{-1}\!\mu \!+\! \mb{W}^{-1}\!
\textstyle \sum_{\nu=1}^R
\bm{\beta}_\nu (\sigma_\nu\tau_\nu)^{-2}\right). \nonumber
\end{align} 
\vspace{-0.3cm}
The linear variance parameter $\tau^2$ follows an inverse-gamma:
\vspace{-0.2cm}
\begin{equation}
\tau_\nu^2 | \mbox{rest} \sim IG((\alpha_\tau + m)/2, (q_\tau +
b_\nu)/2), 
\mbox{\hspace{0.2cm} where \hspace{0.1cm}} b_\nu = (\bm{\beta}_\nu -
\bm{\beta}_0)^\top \mb{W}^{-1} (\bm{\beta}_\nu - \bm{\beta}_0)/\sigma_\nu^2.
\vspace{-0.3cm} \nonumber
\end{equation} 
The linear model covariance matrix $\mb{W}$ follows an inverse-Wishart:
\vspace{-0.3cm}
\begin{equation} 
\mb{W}^{-1}|\mbox{rest} \sim
W_m\left((\rho \mb{V}\mb + \mb{V}_{\hat{W}})^{-1}, \rho+R\right), 
\mbox{\hspace{0.2cm} where \hspace{0.1cm}} \mb{V}_{\hat{W}} = \sum_{\nu=1}^R
\frac{1}{(\sigma_\nu\tau_\nu)^2} (\bm{\beta}_\nu - \bm{\beta}_0)
(\bm{\beta}_\nu - \bm{\beta}_0)^\top.  \nonumber
\vspace{-0.3cm}
\end{equation}
Analytically integrating out $\bm{\beta}_\nu$ and $\sigma_\nu^2$ gives
a marginal posterior for $\mb{K}_\nu$ and improves mixing of the
Markov chain \citep{berg:deol:sans:2001}.  
\vspace{-0.3cm}
\begin{align} 
p(\mb{K}_\nu |\mb{Z}_\nu,\bm{\beta}_0, \mb{W}, \tau^2) = 
& \left(\frac{|\mb{V}_{\tilde{\beta}_\nu}|(2\pi)^{-n_\nu}}{ 
        |\mb{K}_\nu||\mb{W}|\tau^{2m}}\right)^{\frac{1}{2}}
        \frac{\left(\frac{q_\sigma}{2}\right)^{\alpha_\sigma/2}
        \Gamma\left[\frac{1}{2}(\alpha_\sigma + n_\nu)\right]}
        {\left[\frac{1}{2}(q_\sigma+\psi_\nu)\right]^{(\alpha_\sigma+n_\nu)/2}
        \Gamma\left[\frac{\alpha_\sigma}{2}\right]} p(\mb{K}_\nu), 
        \label{e:rj:marginp}
\end{align} 
\vspace{-0.75cm}
\begin{equation}
\mbox{where \hspace{1cm}} \psi_\nu = \mb{Z}_\nu^\top \mb{K}_\nu^{-1} \mb{Z}_\nu +
\bm{\beta}_0^\top \mb{W}^{-1} \bm{\beta}_0/\tau^2 - 
\tilde{\bm{\beta}}_\nu^\top
\mb{V}_{\tilde{\beta}_\nu}^{-1} \tilde{\bm{\beta}}_\nu.
\label{eq:phi}
\end{equation}
Eq.~\eqref{e:rj:marginp} can be used to iteratively obtain draws for
the parameters of $K_\nu(\cdot, \cdot)$ via Metropolis-Hastings (MH),
or as part of the acceptance ratio for proposed modifications to
$\mathcal{T}$ [see Section \ref{sec:tgp:tree}].  
Any hyperparameters to $K_\nu(\cdot, \cdot)$, e.g., parameters to
priors for $\{d,g\}_\nu$ of the isotropic power family, would also
require MH draws.  
The conditional distribution of $\sigma_\nu^2$ with
$\bm{\beta}_\nu$ integrated out allows Gibbs sampling:
\vspace{-0.3cm}
\begin{equation} 
  \sigma_\nu^2 | \mb{Z}_\nu, d_\nu, g, \bm{\beta}_0, \mb{W} \sim 
  IG((\alpha_\sigma + n_\nu)/2, (q_\sigma + \psi_\nu)/2).
  \label{e:s2marginp} 
\vspace{-0.3cm}
\end{equation} 

\vspace{-0.2cm}
\subsubsection{Tree ($\mathcal{T}$)}
\label{sec:tgp:tree}
\vspace{-0.2cm}

Integrating out dependence on the tree structure ($\mathcal{T}$) is
accomplished by RJ-MCMC.  We augment the tree
operations of \citet{chip:geor:mccu:1998}---{\em grow, prune, change,
  swap}---with a rotate operation.  


A {\em change} operation proposes moving an existing split-point $\{u,
s\}$ to either the next greater or lesser value of $s$ ($s_+$ or
$s_-$) along the $u^{\mbox{\tiny th}}$ column of $\mb{X}$.  This is
accomplished by sampling $s'$ uniformly from the set
$\{u_\nu,s_\nu\}_{\nu=1}^{\lceil R/2 \rceil} \times \{+,-\}$ which
causes the MH acceptance ratio for {\em change} to reduce to a simple
likelihood ratio since parameters $\bm{\theta}_r$ in regions $r$ below the
split-point $\{u,s'\}$ are held fixed.

A {\em swap} operation proposes changing the order in which two
adjacent parent-child (internal) nodes split up the inputs.  An
internal parent-child node pair is picked at random from the tree and
their splitting rules are swapped.  However, swaps on parent-child
internal nodes which split on the same variable cause problems because
a child region below both parents becomes empty after the operation.
If instead a {\em rotate} operation from Binary Search Trees (BSTs) is
performed, the proposal will almost always accept.  Rotations are a
way of adjusting the configuration and height of a BST without
violating the BST property, as used, e.g., by {\em red-black
  trees}~\citep{clr}.

\begin{figure}[ht!] 
\begin{center} 
\input{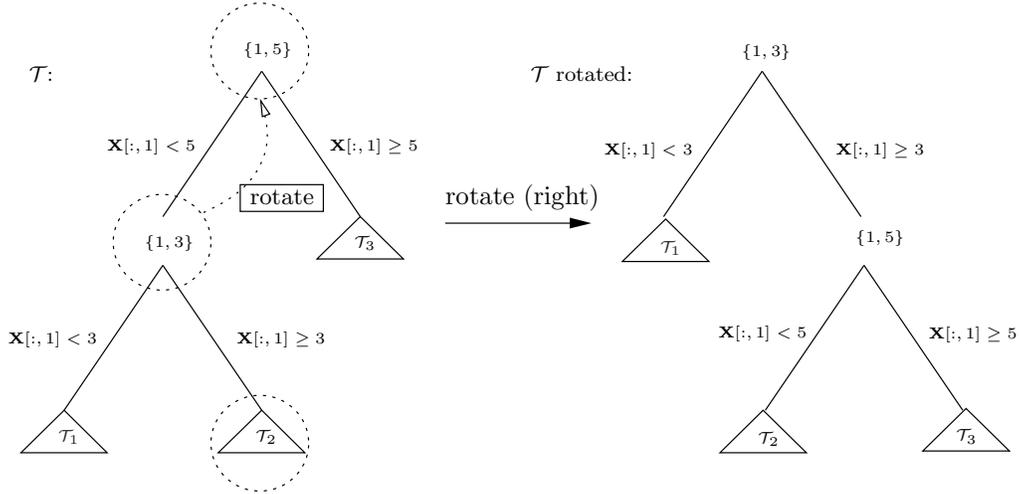} 
\end{center}
\vspace{-0.5cm}
\caption[Example of accepted rotate] {Rotating on the same variable,
  where $\mathcal{T}_1, \mathcal{T}_2, \mathcal{T}_3$ are arbitrary sub-trees 
.} \label{f:rotate} 
\vspace{-0.2cm}
\end{figure}

In the context of a Bayesian MCMC tree proposal, rotations encourage
better mixing of the Markov chain by providing a more dynamic set of
candidate nodes for pruning, thereby helping escape local minima in
the marginal posterior of $\mathcal{T}$.  Figure \ref{f:rotate} shows
an example of a successful right-rotation where a swap would 
produce an empty node (at the current location of $\mathcal{T}_2$).
Since the partitions at the leaves remain unchanged, the likelihood
ratio of a proposed rotate is always 1.  The only ``active'' part of
the MH acceptance ratio is the prior on $\mathcal{T}$, preferring
trees of minimal depth.  Still, calculating the acceptance ratio for a
{\em rotate} is non-trivial because the depth of two of its sub-trees
change.  Figure~\ref{f:rotate} shows a right-rotate, where nodes in
$\mathcal{T}_1$ decrease in depth, while those in $\mathcal{T}_3$
increase.  The opposite is true for left-rotation.  If $I = \{I_i,
I_\ell\}$ is the set of nodes (internals and leaves) of
$\mathcal{T}_1$ and $\mathcal{T}_3$, before rotation, which increase
in depth after rotation, and $D = \{D_i, D_\ell\}$ are those which
decrease in depth, then the MH acceptance ratio for a right-rotate is
\vspace{-0.1cm}
\begin{align*}
\frac{p(\mathcal{T}^*)}{p(\mathcal{T})} =
\frac{p(\mathcal{T}_1^*) p(\mathcal{T}_3^*)}
        {p(\mathcal{T}_1) p(\mathcal{T}_3)} 
&= \frac{\prod_{\eta\in I_i} a(2+q_\eta)^{-b} 
  \prod_{\eta\in I_\ell} [1-a(2+q_\eta)^{-b}]}
{\prod_{\eta\in I_i} a(1+q_\eta)^{-b} 
  \prod_{\eta\in I_\ell} 
        [1-a(1+q_\eta)^{-b}]} \\
      &\times \frac{\prod_{\eta\in D_i} a q_\eta^{-b} 
  \prod_{\eta\in D_\ell} [1-a q_\eta^{-b}]}
{\prod_{\eta\in D_i} a(1+q_\eta)^{-b}
  \prod_{\eta\in D_\ell} [1-a(1+q_\eta)^{-b}]}.
\vspace{-0.2cm}
\end{align*}
The MH acceptance ratio for a left-rotate is analogous.

{\em Grow} and {\em prune} operations are complex because they add or
remove partitions, changing the dimension of the parameter space.  The
first step for either operation is to uniformly select a leaf node
(for {\em grow}), or the parent of a pair of leaf nodes (for {\em
  prune}). When a new region $r$ is added, new parameters
$\{K(\cdot,\cdot), \tau^2\}_r$ must be proposed, and when a region is
taken away the parameters must be absorbed by the parent region, or
discarded. When evaluating the MH acceptance ratio the linear model
parameters $\{\bm{\beta}, \sigma^2\}_r$ are integrated out
\eqref{e:rj:marginp}.  One of the newly grown children is uniformly
chosen to receive the correlation function $K(\cdot, \cdot)$ of its
parent, essentially inheriting a block from its parent's correlation
matrix. To ensure that the resulting Markov chain is ergodic and
reversible, the other new sibling draws its $K(\cdot,\cdot)$ from the
prior thus giving a unity Jacobian term in the RJ-MCMC.  Note that
{\em grow} operations are the only place where priors are used as
proposals; random-walk proposals are used elsewhere [see Section
\ref{sec:tgp:implement}].

Symmetrically, {\em prune} operations randomly select parameters from
$K(\cdot,\cdot)$ for the consolidated node from one of the children
being absorbed.  After accepting a {\em grow} or {\em prune},
$\sigma^2_r$ can be drawn from its marginal posterior, with
$\bm{\beta}_r$ integrated out (\ref{e:s2marginp}), followed by draws
for $\bm{\beta}_r$ and the rest of the parameters in the
$r^{\mbox{\tiny th}}$ region.

Let $\{\mb{X}, \mb{Z}\}$ be the data at the new parent node $\eta$ at
depth $q_\eta$, and $\{\mb{X}_1, \mb{Z}_1\}$ and $\{\mb{X}_2,
\mb{Z}_2\}$ be the partitioned child data at depth $q_\eta+1$ created
by a new split $\{u,s\}$.  Also, let $\mathcal{P}$ be the set of
prunable nodes of $\mathcal{T}$, and $\mathcal{G}$ the set of growable
nodes.  If $\mathcal{P'}$ are the prunable nodes in
$\mathcal{T'}$---after the (successful) {\em grow} at $\eta$---and the
parent of $\eta$ was prunable in $\mathcal{T}$, then
$|\mathcal{P}'|=|\mathcal{P}|$.  Otherwise
$|\mathcal{P}'|=|\mathcal{P}|+1$.  The MH ratio for {\em grow} is:
\vspace{-0.1cm}
\begin{equation}
\frac{|\mathcal{G}|}{|\mathcal{P}'|} 
\frac{a(1+q_\eta)^{-b} (1-a(2+q_\eta)^{-b})^2} {1-a(1+q_\eta)^{-b}} 
\frac{p(\mb{K}_1, |\mb{Z}_1,\bm{\beta}_0, \tau^2_1, \mb{W}) 
  p(\mb{K}_2|\mb{Z}_2,\bm{\beta}_0, \tau^2_2, \mb{W})}
{p(\mb{K} | \mb{Z},\bm{\beta}_0, \tau^2, \mb{W}) q(\mb{K}_2)}
\vspace{-0.1cm}
\end{equation}
assuming that $\mb{K}_1$ was randomly chosen to receive the
parameterization of its parent, $\mb{K}$, and that the new parameters
for $\mb{K}_2$ are proposed according to $q$.  The {\em prune}
operation is analogous.  Note that for
the posteriors $p(\mb{K}|\mb{Z},\bm{\beta}_0, \tau^2, \mb{W})$,\\
$p(\mb{K}_1 |\mb{Z}_1,\bm{\beta}_0, \tau^2_1, \mb{W})$ and
$p(\mb{K}_2|\mb{Z}_2,\bm{\beta}_0, \tau^2_2, \mb{W})$, the
``constant'' terms in (\ref{e:rj:marginp}) are required because they
do not occur the same number of times in the numerator and
denominator.

\vspace{-0.4cm}
\subsection{Treed GP Prediction}
\label{sec:tgp:pred}
\vspace{-0.3cm}

Prediction under the above GP model is straightforward
\citep{hjor:omre:1994}.  Conditional on the covariance structure, the
predicted value of $z(\mb{x}\in r_\nu$) is normally distributed with
mean and variance \vspace{-0.2cm}
\begin{align} 
 \hat{z}(\mb{x}) &= E(\mb{Z}(\mb{x}) | 
        \; \mbox{data}, \mb{x}\in r_\nu ) \; = \;
 \mb{f}^\top(\mb{x}) \tilde{\bm{\beta}}_\nu +
        \mb{k}_\nu(\mb{x})^\top \mb{K}_\nu^{-1}(\mb{Z}_\nu -
        \mb{F}_\nu\tilde{\bm{\beta}}_\nu), \label{eq:predmean} \\ 
 \hat{\sigma}(\mb{x})^2 &= 
        \mbox{Var}(\mb{z}(\mb{x}) | \;\mbox{data}, \mb{x}\in r_\nu ) 
  \; = \; \sigma_\nu^2 [\kappa(\mb{x}, \mb{x}) - 
        \mb{q}_\nu^\top(\mb{x})\mb{C}_\nu^{-1} \mb{q}_\nu(\mb{x})],
        \label{eq:predvar} 
\end{align}
\vspace{-1.5cm}
\begin{align}
\mbox{where} &&
\mb{C}_\nu^{-1} &= (\mb{K}_\nu + 
        \tau_\nu^2\mb{F}_\nu \mb{W} \mb{F}_\nu^\top)^{-1} &
\mb{q}_\nu(\mb{x}) &= 
        \mb{k}_\nu(\mb{x}) + \tau_\nu^2\mb{F}_\nu \mb{W}
        \mb{f}(\mb{x}) \label{eq:auxpred} \\ 
&& \kappa(\mb{x},\mb{y}) &= K_\nu(\mb{x},\mb{y}) 
        + \tau_\nu^2\mb{f}^\top(\mb{x}) \mb{W} \mb{f}(\mb{y}) \nonumber
\end{align} 
with $\mb{f}^\top(\mb{x}) = (1, \mb{x}^\top)$, and
$\mb{k}_\nu(\mb{x})$ is a $n_\nu-$vector with $\mb{k}_{\nu,j}(\mb{x})=
K_\nu(\mb{x}, \mb{x}_j)$, for all $\mb{x}_j \in \mb{X}_\nu$.  

Conditional on a particular tree, $\mathcal{T}$, the posterior
predictive surface described in
Eqs.~(\ref{eq:predmean}--\ref{eq:predvar}) is discontinuous across the
partition boundaries of $\mathcal{T}$.  However, in the aggregate of
samples collected from the joint posterior distribution of
$(\mathcal{T}, \bm{\theta})$, the mean tends to smooth out near
likely partition boundaries as the tree operations {\em grow, prune,
  change}, and {\em swap} integrate over trees and GPs with larger
posterior probability.  Uncertainty in the posterior for $\mathcal{T}$
translates into higher posterior predictive uncertainty near region
boundaries.  When the data actually indicate a non-smooth process, the
treed GP retains the flexibility necessary to model discontinuities.
When the data are consistent with a continuous process, as in the
motorcycle data example in Section~\ref{sec:tgp:results}, the treed GP
fits are almost indistinguishable from continuous.

\vspace{-0.4cm}
\subsection{Implementation}
\label{sec:tgp:implement}
\vspace{-0.3cm}

The treed GP model is coded in a mixture of {\tt C} and {\tt C++},
using {\tt C++} for the tree structure and {\tt C} for the GP at each
leaf of $\mathcal{T}$.  The {\tt C} code can interface with either
standard platform-specific {\tt Fortran} {\tt BLAS/Lapack} libraries
for the necessary linear algebra routines, or link to those
automatically configured for fast execution on a variety of platforms
via the {\tt ATLAS} library \citep{atlas-hp}.  To improve usability,
the routines have been wrapped up in an intuitive {\tt R} interface,
and are available on CRAN \citep{cran:R} at
\begin{center}
\verb!http://www.cran.r-project.org/web/packages/tgp/index.html!
\end{center}
as a package called {\tt tgp}.

It is useful to first translate and re-scale the input dataset ($\mb{X}$)
so that it lies in an $\Re^{m_X}$ dimensional unit cube. This makes it
easier to construct prior distributions for the width parameters to
the correlation function $K(\cdot,\cdot)$.  Conditioning on
$\mathcal{T}$, proposals for all parameters which require MH sampling
are taken from a uniform ``sliding window'' centered around the
location of the last accepted setting.  For example, a proposed a new
nugget parameter $g_\nu$ to the correlation function $K(\cdot, \cdot)$
in region $r_\nu$ would go as $g_\nu^* \sim \mbox{Unif}\left(3g_\nu/4,
  4g_\nu/3 \right)$.  Calculating the forward and backward proposal
probabilities for the MH acceptance ratio is straightforward.

After conditioning on $(\mathcal{T}, \bm{\theta})$, prediction can
be parallelized by using a producer-consumer model.  This allows the
use of {\tt PThreads} in order to take advantage of multiple
processors, and get speed-ups of at least a factor of two, which is
helpful as multi-processor machines become commonplace.  Parallel
sampling of the posterior of $\bm{\theta}|\mathcal{T}$ for each of the
$\{\theta_\nu\}_{\nu=1}^R$ is also possible.

\vspace{-0.2cm}
\subsection{Illustration}
\label{sec:tgp:results}
\vspace{-0.1cm}

In this section we illustrate the treed GP model on the Motorcycle
Accident Dataset \citep{silv:1985}, a classic dataset used in recent
literature \citep[e.g.,][]{rasm:ghah:nips:2002} to demonstrate the
success of nonstationary models.  The dataset consists of measurements
of the acceleration of the head of a motorcycle rider, which we attempt to
predict as a function of
time in the first moments after an impact.  In addition to suggesting
a model with a nonstationary covariance structure, there is
input-dependent noise (a.k.a., 
heteroscedasticity).  To keep things simple in this illustration, the
isotropic power family (\ref{eq:pow}) correlation function ($p_0=2$)
is chosen for $K^*(\cdot,\cdot|d)$ with range parameter $d$, combined
with nugget $g$ to form $K(\cdot, \cdot|d,g)$.

\begin{figure}[ht!] 
\begin{center}
\includegraphics[scale=0.45,trim=50 40 25 25]{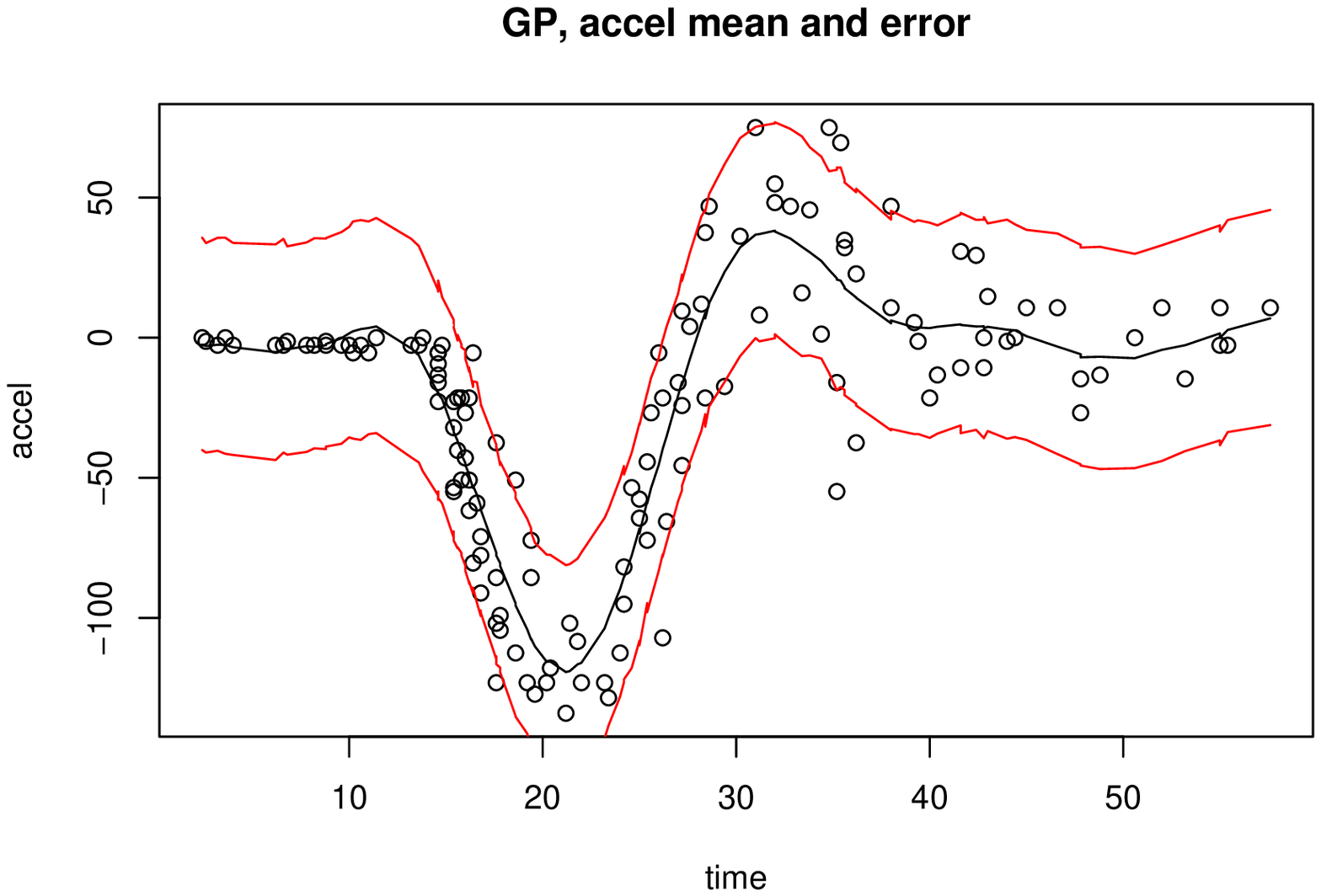} 
\includegraphics[scale=0.45,trim=0 40 50 25]{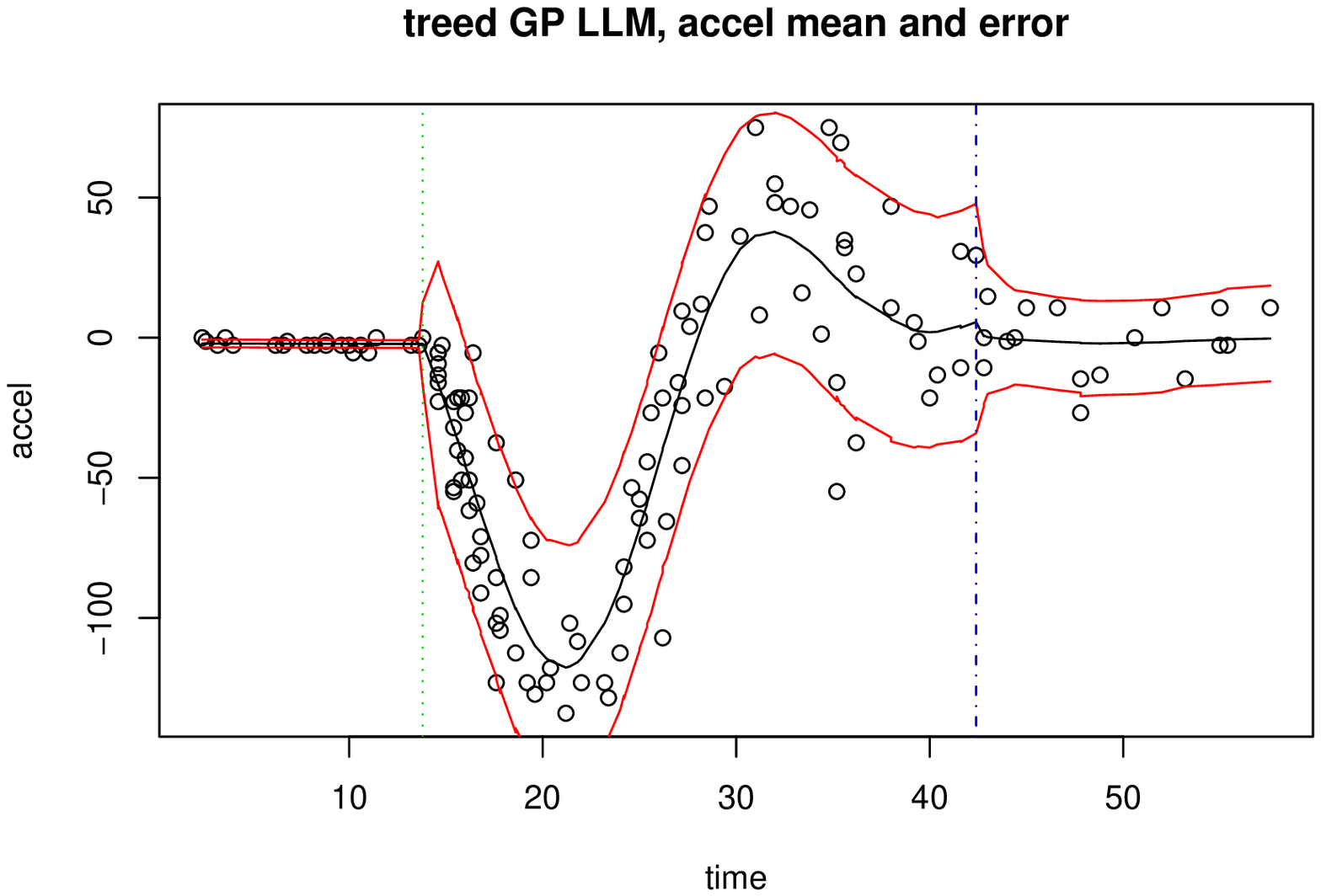} 
\end{center} 
\vspace{-0.3cm}
\caption[Treed GP results on Motorcycle Accident Data]
{Motorcycle Dataset, fit ({\em left}) by a stationary process and 
{\em right} by our nonstationary model.} \label{f:moto}
\vspace{-0.2cm}
\end{figure} 

Figure~\ref{f:moto} shows the data and the fits given by both a
stationary GP (left) and the treed GP model (right), along with 90\%
credible intervals.  For the treed GP, vertical lines illustrate a
typical treed partition $\mathcal{T}$.  Notice that the stationary GP
is completely unable to capture the heteroscedasticity, and that the
large variability in the central region drives both ends to be more
wiggly (in particular, the transition from the flat left initial
region requires an upward curve before descending).  In contrast, the
treed GP clearly reflects the differing levels of uncertainty, as well
as allowing a flatter fit to the initial segment and a smoother fit to
the final segment.  20,000 MCMC rounds yielded an average of 3.11
partitions in $\mathcal{T}$.

These results differ from those of Rasmussen \& Ghahramani (2002).  In
particular, the error-bars they report for the leftmost region seem
too large relative to the other regions.  They use what they call an
``infinite mixture of GP experts'' which is a Dirichlet
process mixture of GPs.  They report that the posterior distribution
uses between 3 and 10 experts to fit this dataset, with even 10-15
experts still having considerable posterior mass, although 
there are ``roughly'' three regions.  Contrast this with the treed GP
model which almost always partitions into three regions, occasionally
four, rarely two.  On speed grounds, the treed GP is also a winner.
Running the mixture of GP experts model using a total of 11,000 MCMC
rounds, discarding the first 1,000, took roughly one hour on a 1 GHz
Pentium.  Allowing treed GP to use 25,000 MCMC rounds, discarding the
first 5,000, takes about 3 minutes on a 1.8 GHz Athalon.

\begin{figure}[ht!] 
\centering
\vspace{-0.75cm}
\includegraphics[scale=0.6,angle=-90,trim=0 0 74 30,clip=TRUE]{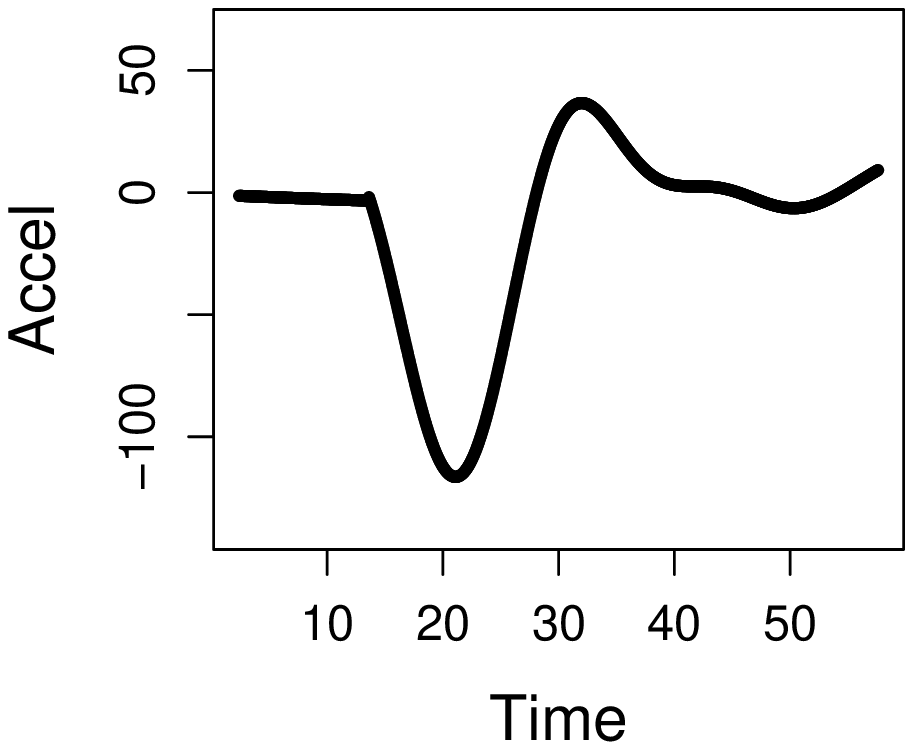}
\includegraphics[scale=0.6,angle=-90,trim=0 58 74 30,clip=TRUE]{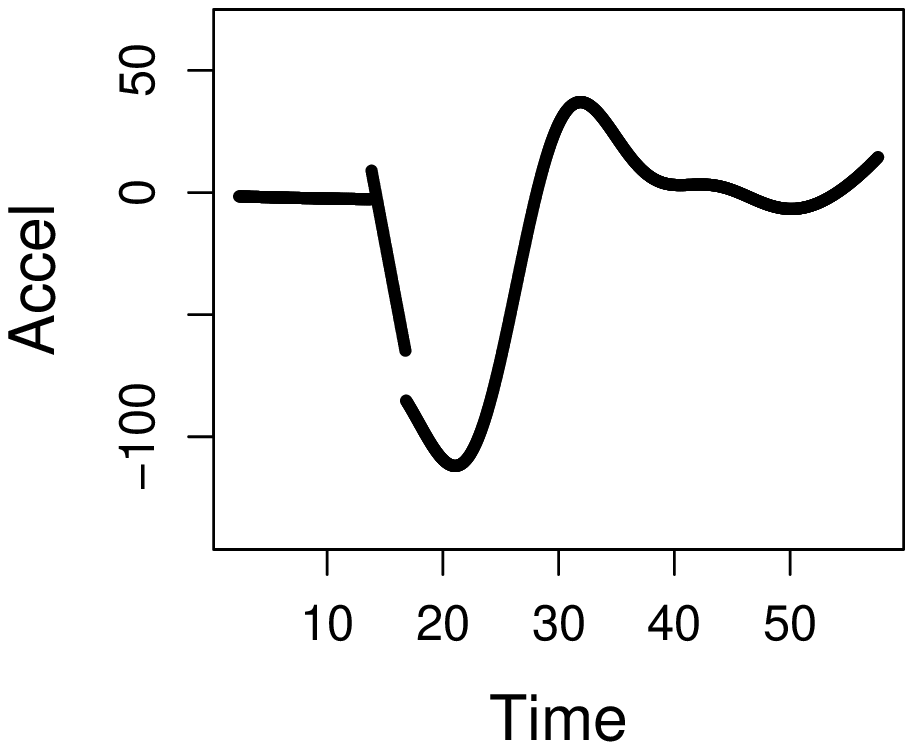}
\includegraphics[scale=0.6,angle=-90,trim=0 58 74 30,clip=TRUE]{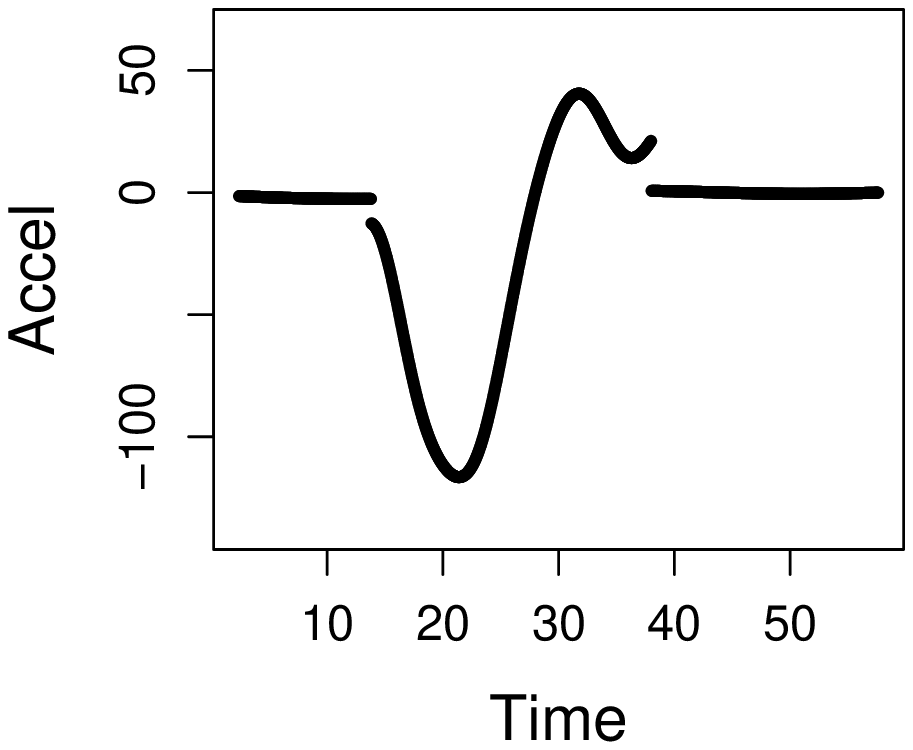}
\includegraphics[scale=0.6,angle=-90,trim=53 0 0 30,clip=TRUE]{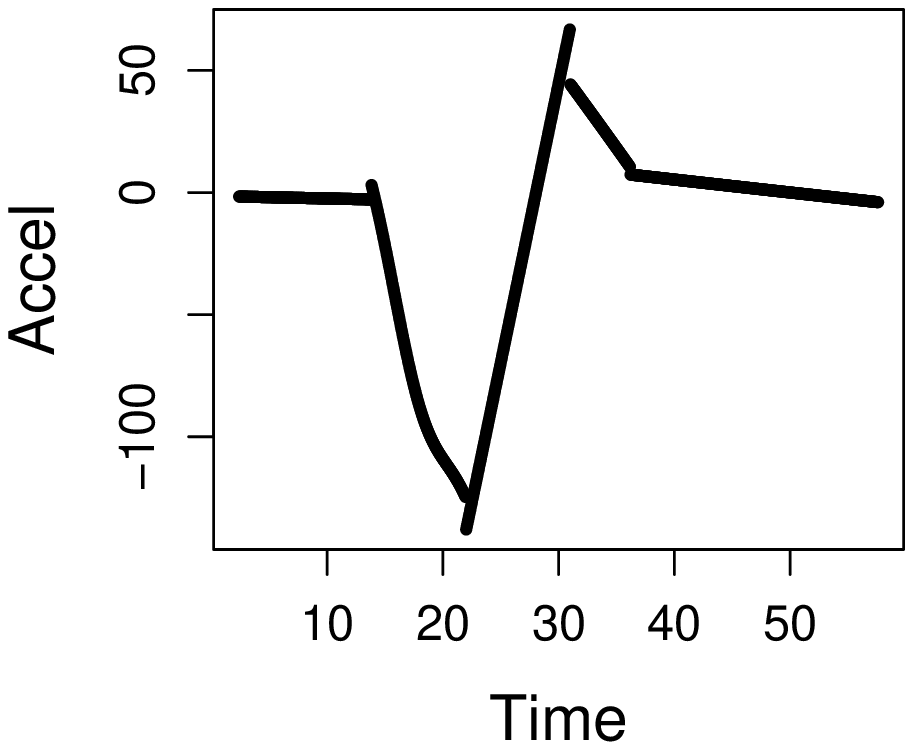}
\includegraphics[scale=0.6,angle=-90,trim=53 58 0 30,clip=TRUE]{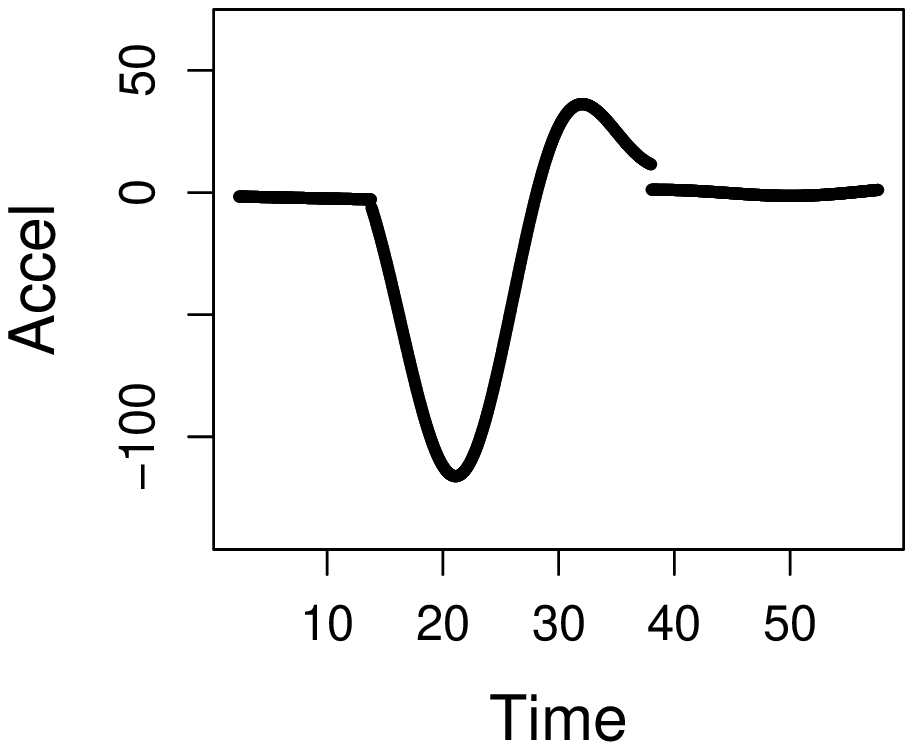}
\includegraphics[scale=0.6,angle=-90,trim=53 58 0 30,clip=TRUE]{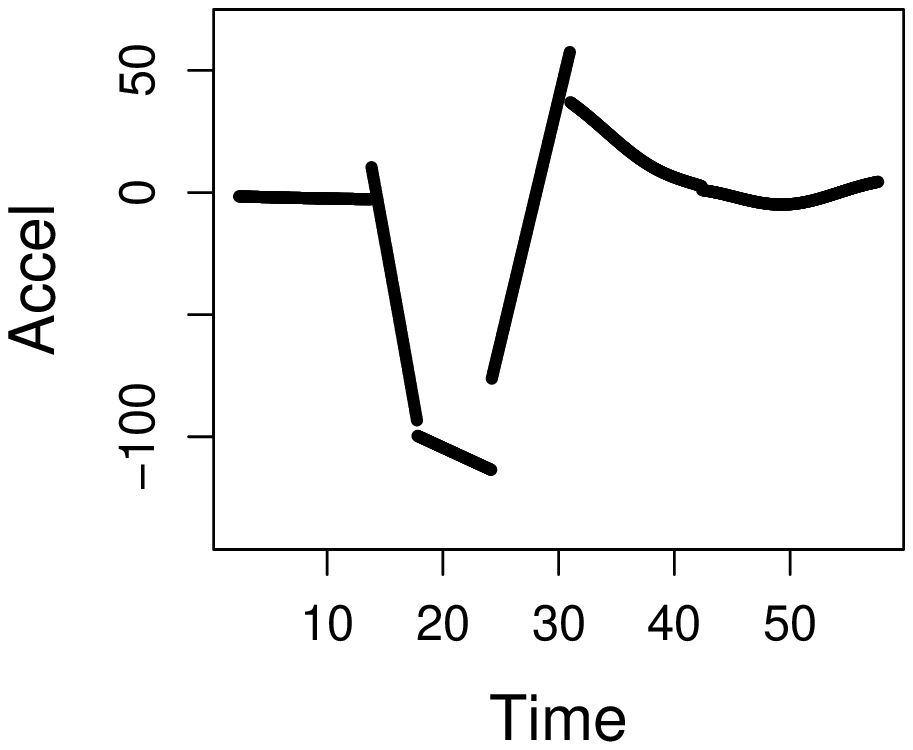}
\vspace{-0.3cm}
\caption{Fits of the motorcycle accident data from individual MCMC
  iterations.} \label{f:moto-samples}
\end{figure}

We note that the mean fitted function in the right plot in
Figure~\ref{f:moto} is essentially that of a continuous function.
Figure~\ref{f:moto-samples} shows examples of the fits from individual
MCMC iterations that are eventually averaged.  While the individual
partition models are typically discontinuous, it is clear from
Figure~\ref{f:moto} that the mean fitted function is well-behaved.

\vspace{-0.2cm}
\subsection{Limiting linear models}
\label{sec:gpllm}
\vspace{-0.1cm}

In some cases, a GP may not be needed within a
partition, and a much simpler model, such as a linear model, may
suffice.  In particular, because of the linear mean function in our
implementation of the GP, the standard linear model can be seen as a
limiting case.  The linear model is then more parsimonious, as well as
much more computationally efficient.  Use of a model-switching prior
allows practical implementation.  More details are available in
\citet{gra:lee:2008b}.  The value of such an approach can be seen from
the fit on the right side of Figure~\ref{f:moto}.  The leftmost
partition looks quite flat, and so could be fit just as well with a
line rather than a GP.  The center partition clearly requires a GP
fit.  The rightmost partition looks mostly linear, and would give a
posterior which is a mix of a GP and linear model.  Indeed,
Figure~\ref{f:moto-samples} shows examples of the individual fits, and
the leftmost section is nearly always flat, the rightmost section is
often but not always flat, and the center section is typically curved
but even there it can be essentially piecewise linear (the range
parameter $d$ is estimated to be large, giving a nearly linear fit).
Replacing the full
GP with a linear model in a partition greatly reduces the
computational resources required to update the model in that partition.
Treed and non-treed Gaussian process with jumps to the limiting linear
model are implemented in the {\tt tgp} package on CRAN, and we take
advantage of the full formulation in our analyses herein.

\vspace{-0.3cm}
\section{Rocket Booster Model Results}
\label{sec:lgbb:results}
\vspace{-0.2cm}

We fit our treed GP model to the rocket booster data using ten
independent RJ-MCMC chains with 15,000 MCMC rounds each.  The first
5,000 were discarded as burn-in, and every tenth thereafter was treated
as a sample from the posterior distribution
$\pi(\mathcal{T},\bm{\theta}|Z)$.  In total, 10,000 samples were
saved.
\begin{figure}[ht!]
\begin{center}
\includegraphics[angle=-90,scale=0.33,trim=20 185 50 175]{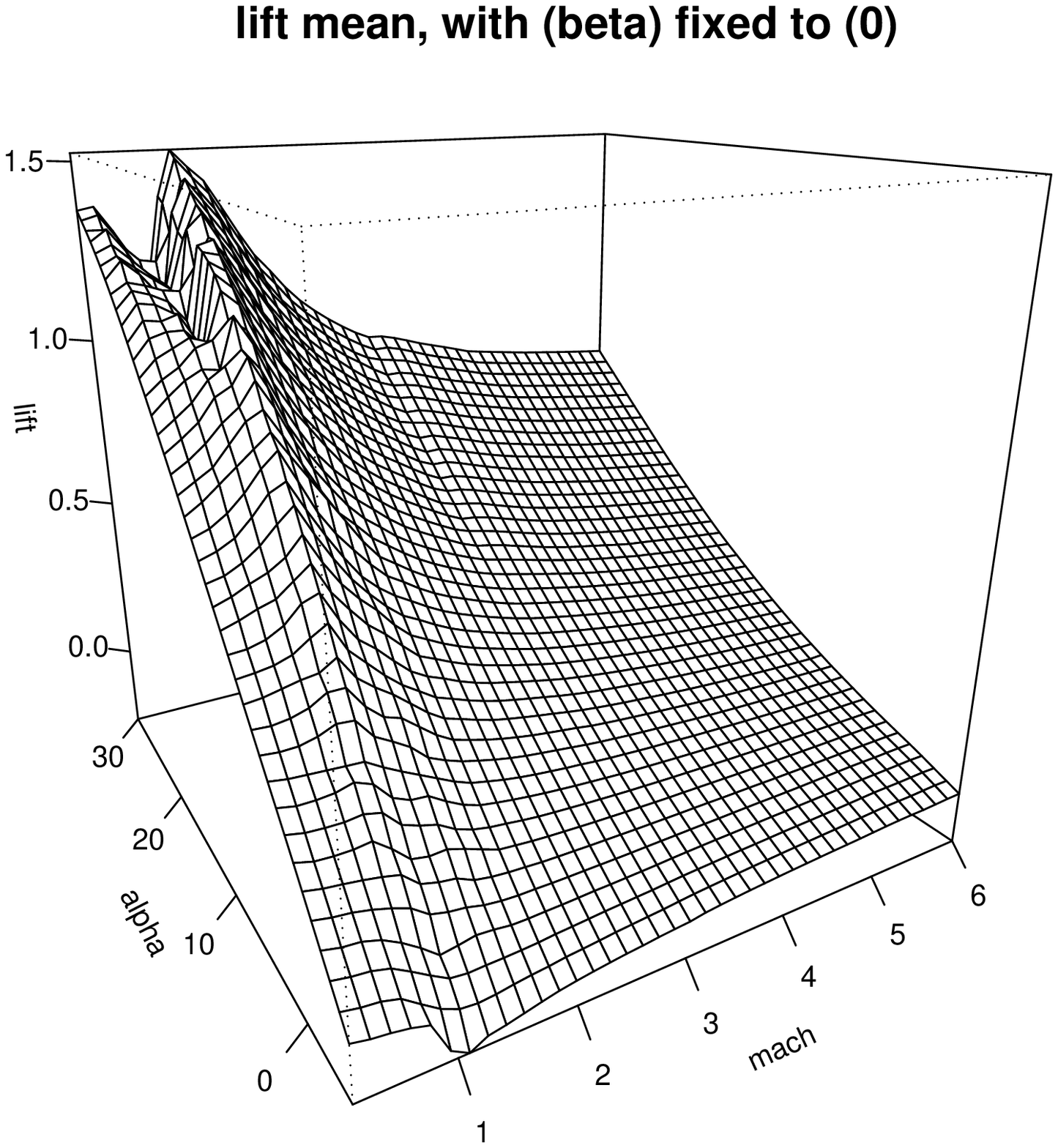} 
\includegraphics[angle=-90,scale=0.33,trim=20 185 50 175]{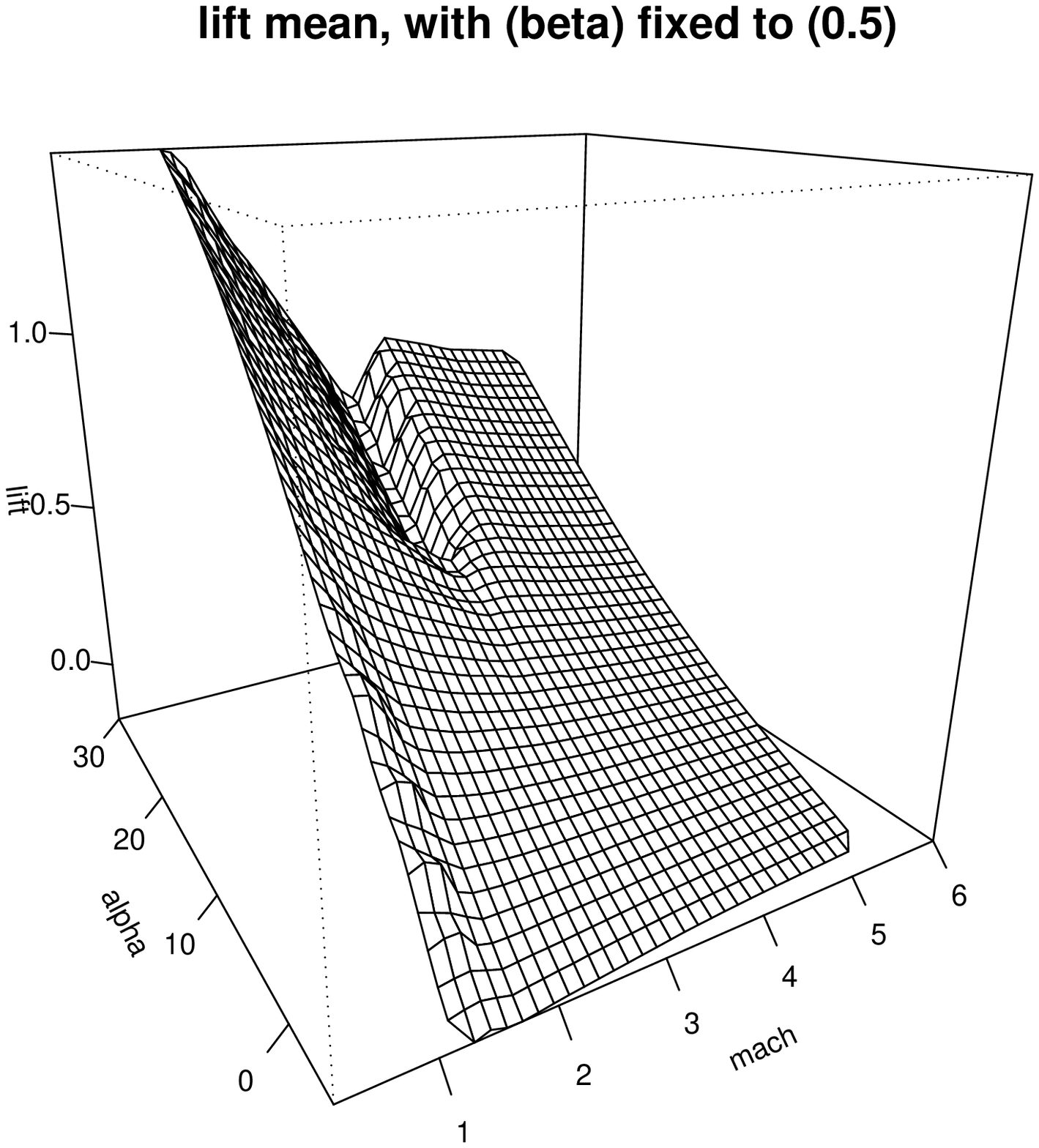}
\includegraphics[angle=-90,scale=0.33,trim=20 185 50 185]{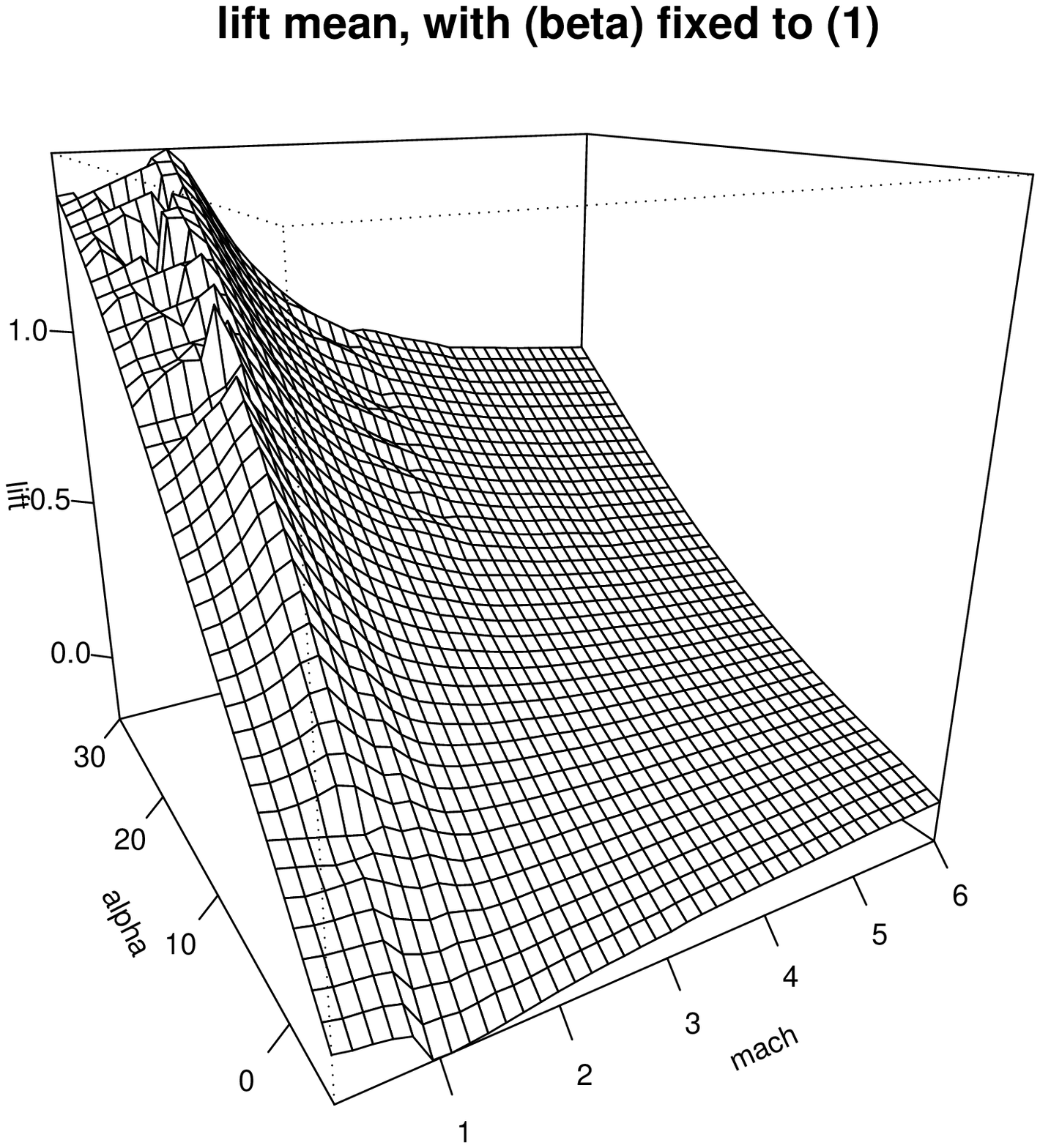}
\includegraphics[angle=-90,scale=0.33,trim=20 185 75 175]{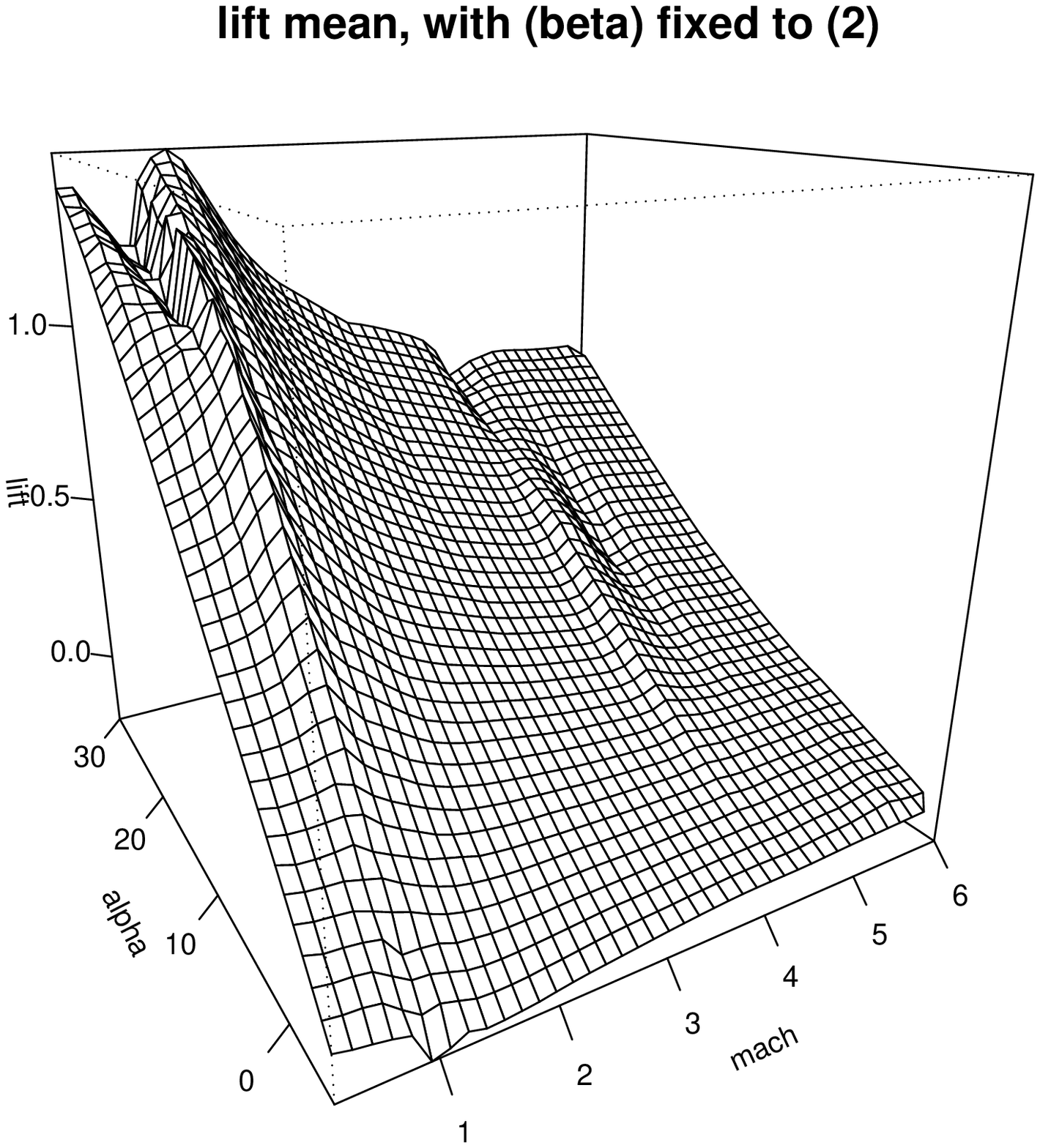}
\includegraphics[angle=-90,scale=0.33,trim=20 185 75 175]{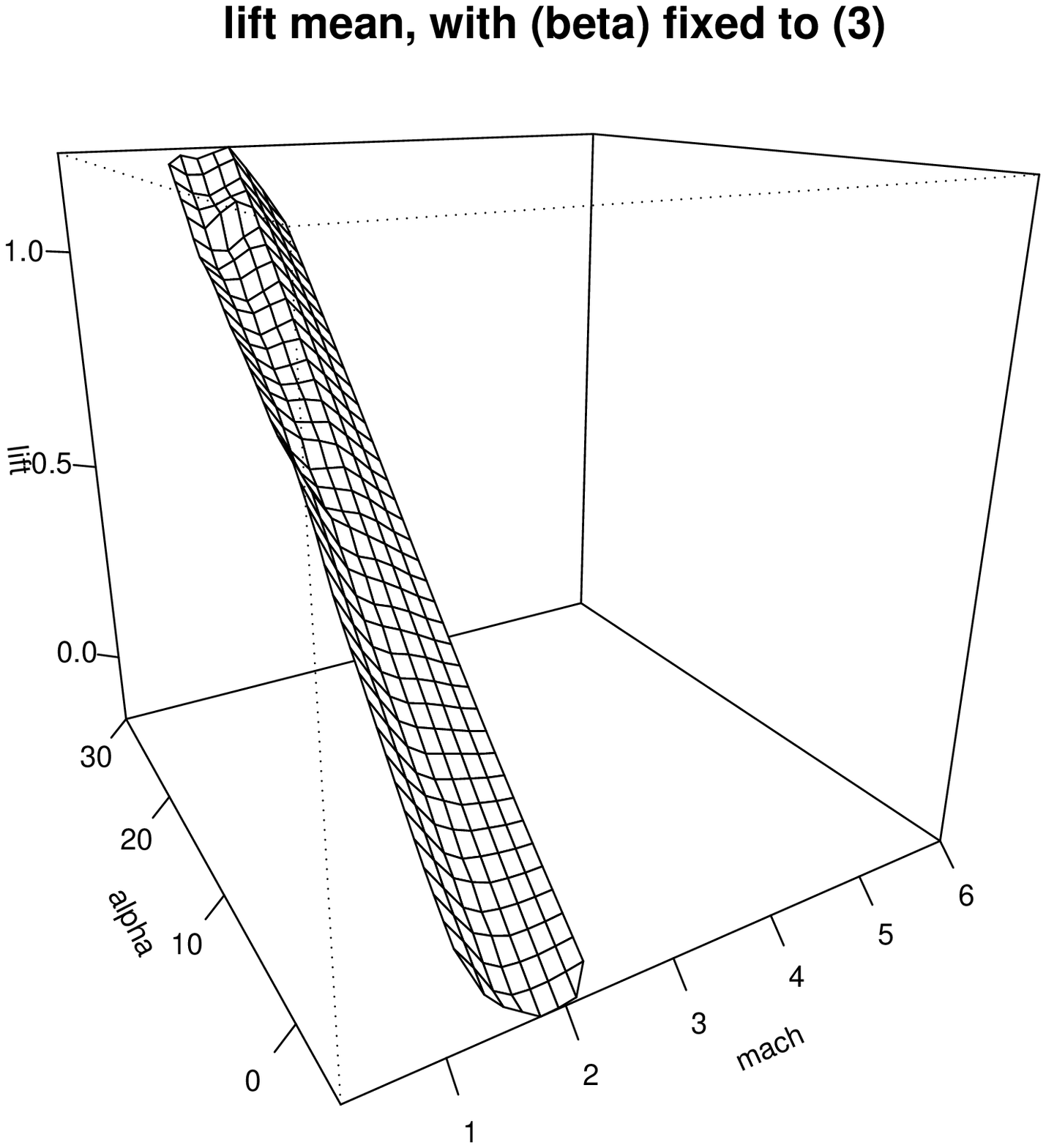}
\includegraphics[angle=-90,scale=0.33,trim=20 185 75 185]{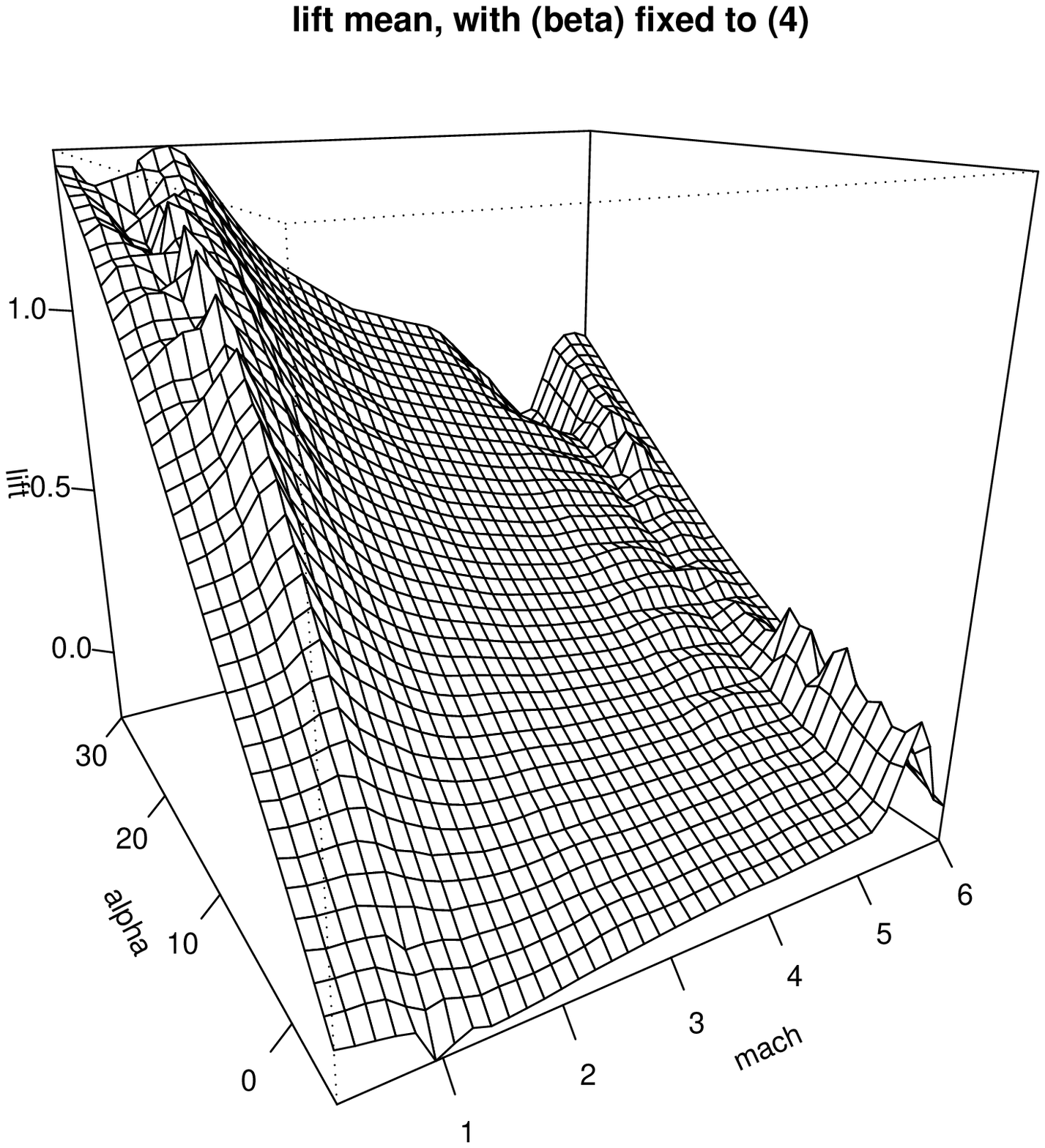} 
\end{center}
\vspace{-0.3cm}
\caption{Posterior predictive mean surfaces of lift for all sideslip angles.
  Note that for levels 0.5 and 3 (center), Mach ranges only in $(1,5)$
  and $(1.2,2.2)$}
\label{f:lgbb:surf}
\vspace{-0.2cm}
\end{figure}
On a single 3.2 GHz Xeon processor this took about 60 hours.  On the
same machine, using the same (tuned) linear algebra libraries,
inverting a single $3041 \times 3041$ matrix takes about 17 seconds,
so obtaining the same number of samples from a stationary GP would
have taken a minimum of 708 hours.  This is a gross underestimate
because it assumes only one inverse is needed per MCMC round.
Moreover, it does not count any of the $O(n^2)$ operations like
determinants of $\mb{K}$ (assuming a factorized $\mb{K}$ was saved in
computing $\mb{K}^{-1}$) or multiplications like
$\mb{Z}\mb{K}^{-1}\mb{Z}$ in (\ref{eq:phi}), nor does it factor in the
time needed to sample from the posterior predictive distribution.


\begin{figure}
\begin{center}
\includegraphics[angle=-90,scale=0.19,trim=0 100 20 0]{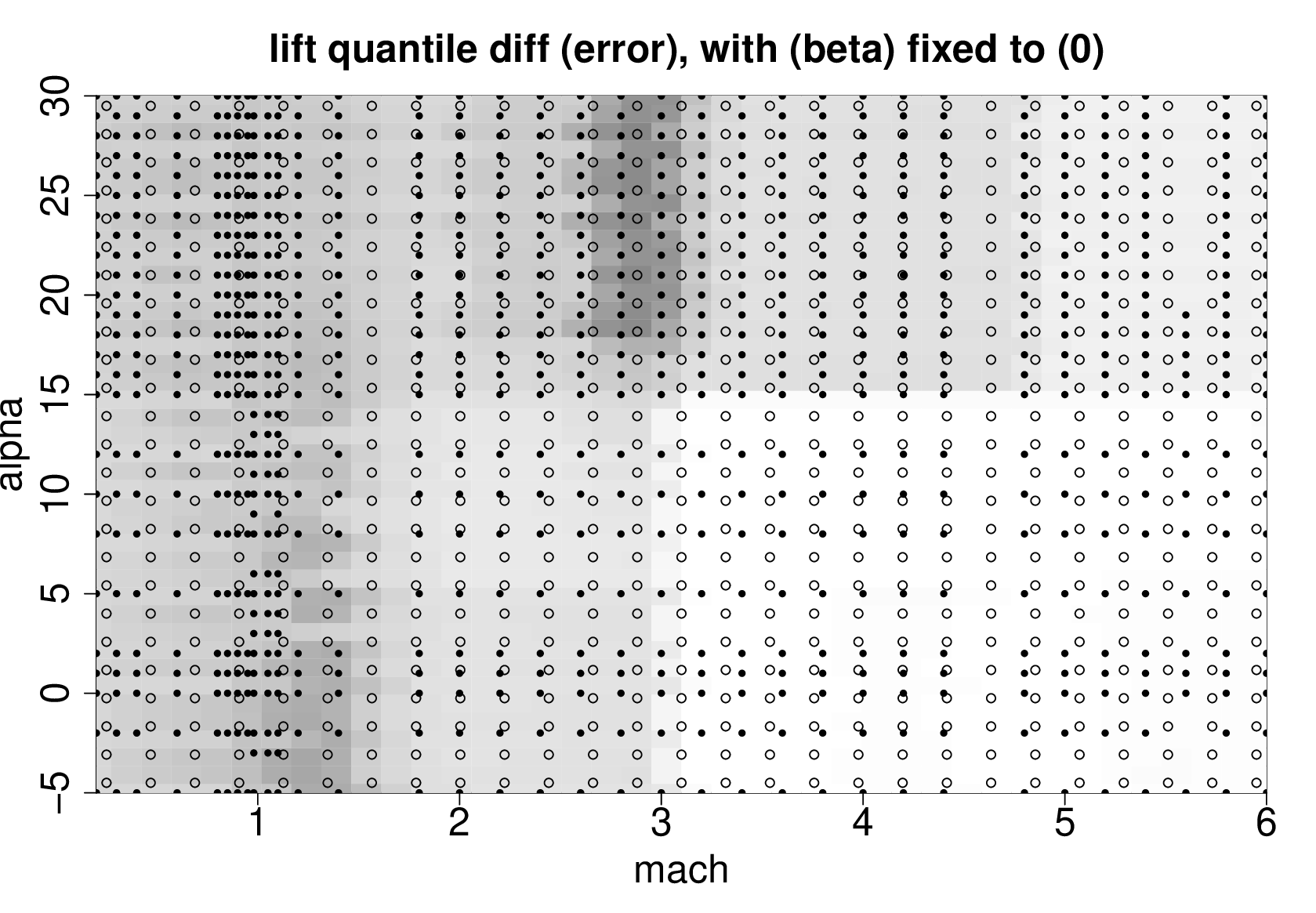} 
\includegraphics[angle=-90,scale=0.19,trim=0 30 20 0]{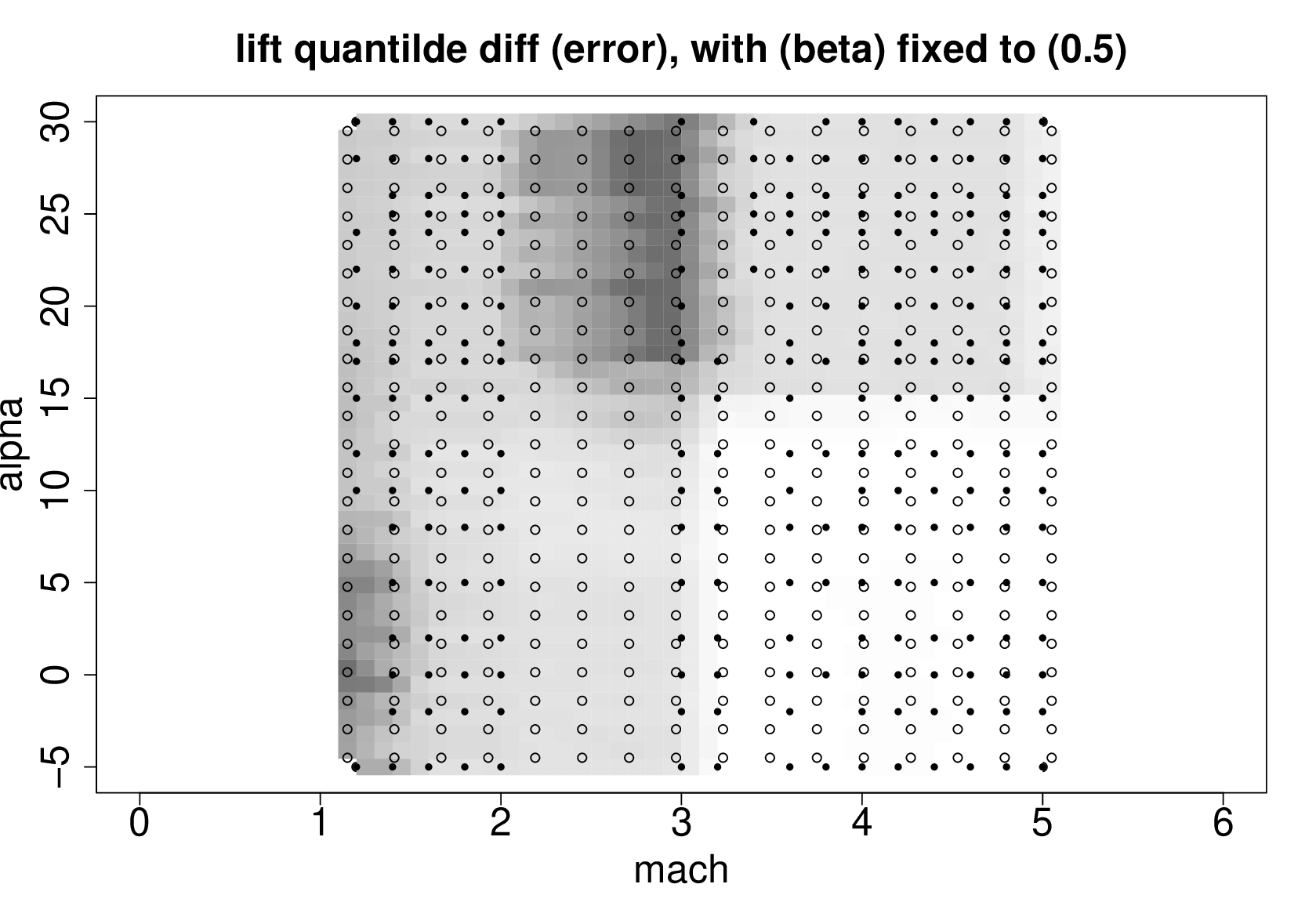}
\includegraphics[angle=-90,scale=0.19,trim=0 30 20 100]{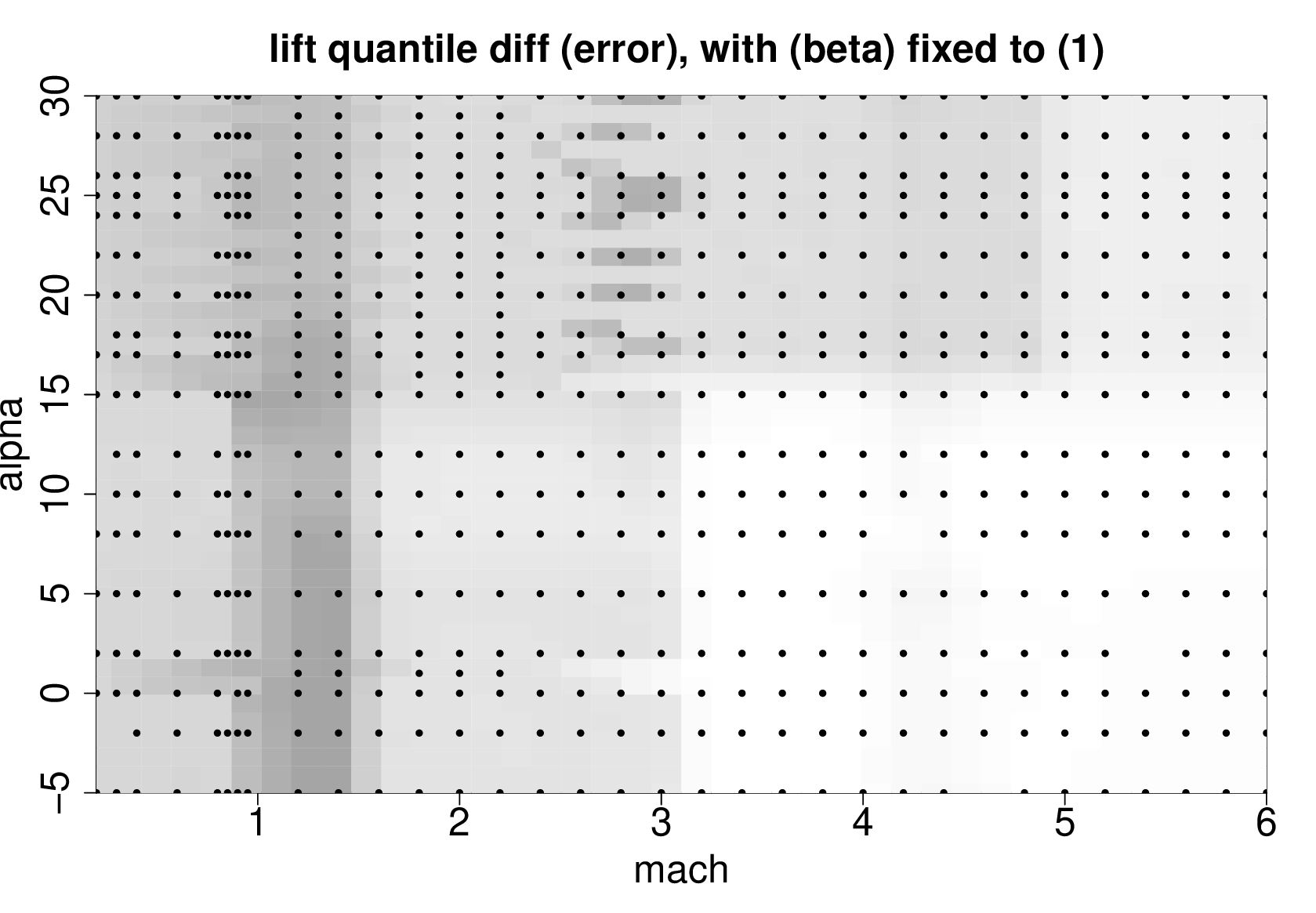}
\includegraphics[angle=-90,scale=0.19,trim=0 100 50 0]{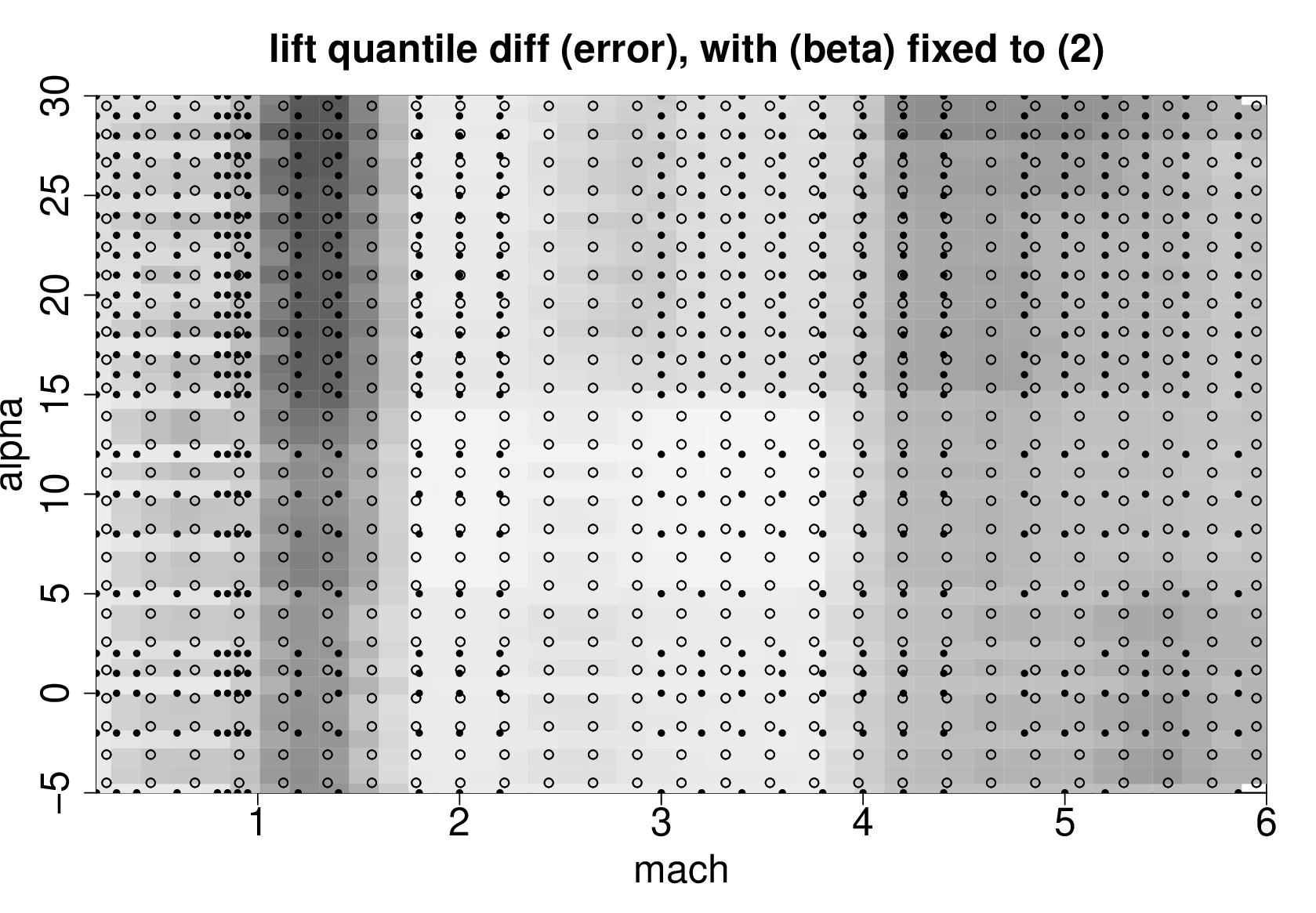}
\includegraphics[angle=-90,scale=0.19,trim=0 30 50 0]{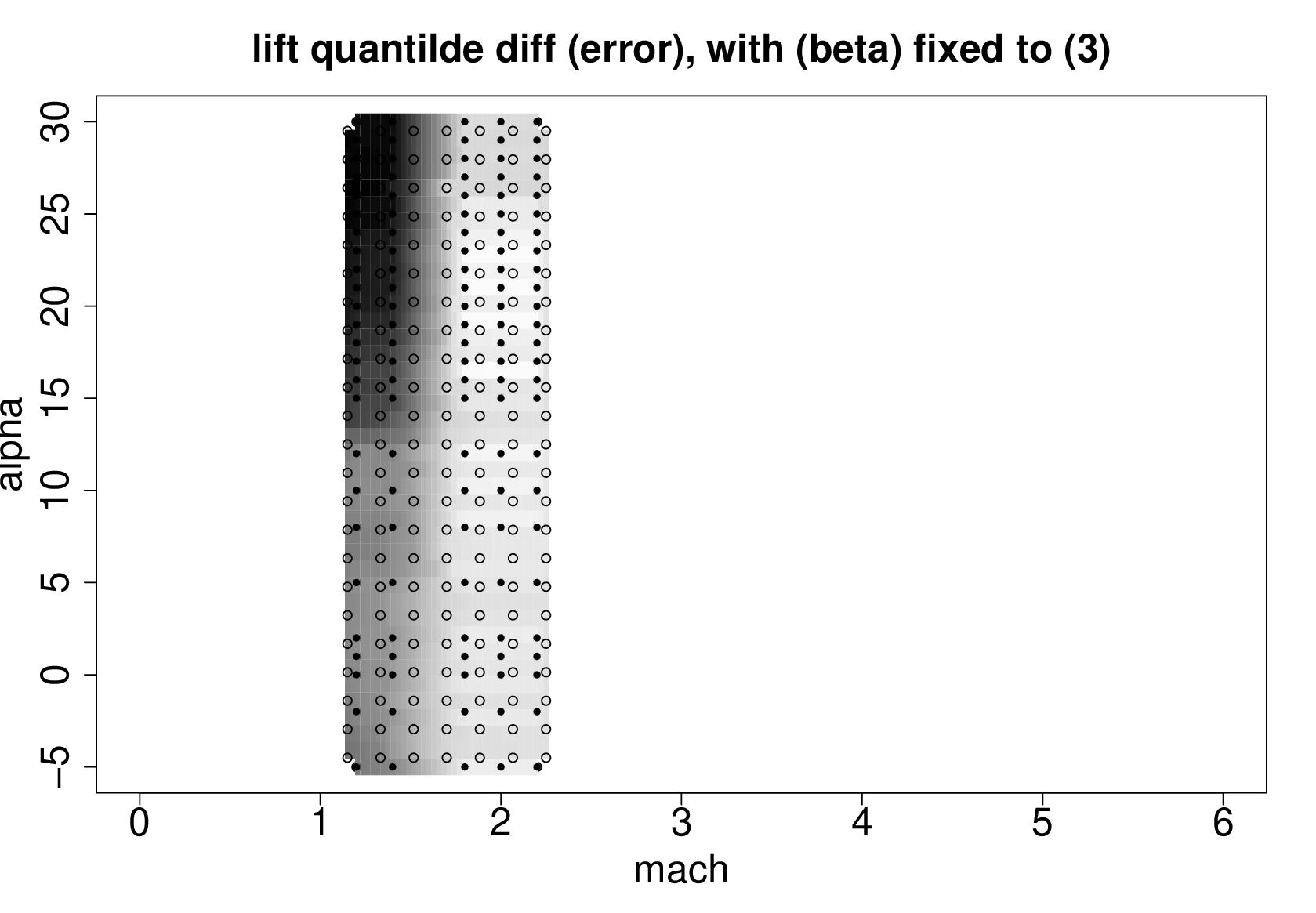}
\includegraphics[angle=-90,scale=0.19,trim=0 30 50 100]{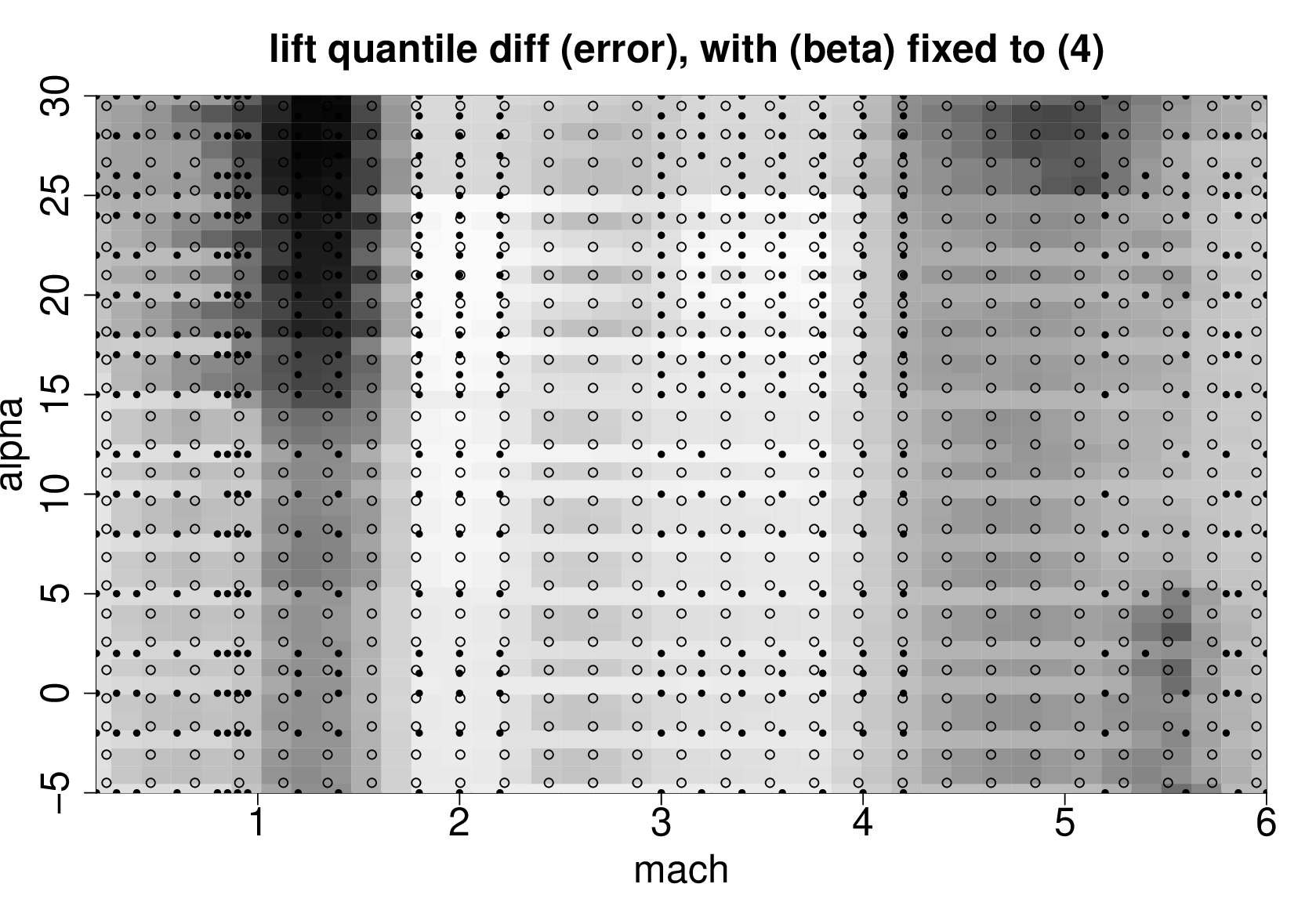} 
\end{center}
\vspace{-0.2cm}
\caption{Posterior predictive variance surfaces of lift
  for all six sideslip angles.  Dots show the locations of
  experimental runs.  Darker shades are higher values.
}
\label{f:lgbb:var}
\vspace{-0.2cm}
\end{figure}

Figures~\ref{f:lgbb:surf} \& \ref{f:lgbb:var} summarize the posterior
predictive distribution for the lift response for each of the six
levels of sideslip angle.  Figure~\ref{f:lgbb:surf} contains plots of
the fitted mean lift surface by speed and angle of attack, and
Figure~\ref{f:lgbb:var} plots a measure of the estimated predictive
uncertainty given by the difference in 95\% and 5\% quantiles of
samples from the posterior predictive
distribution.  
The treed GP works well here.  Most of the space is
nicely smooth, with the sharp transition at Mach one also
well-modeled.  Most of the potential false convergences have been
smoothed out.  But the estimated variability reflects both increased
variability where the function is changing rapidly (e.g., near Mach
one, particularly for higher sideslip levels) and especially 
where there are issues of possible false numerical convergence.  Note
that the uncertainty is not that high near Mach one at sideslip level
zero because of the large number of samples taken in that region.  We
also note that the increased uncertainty seen in the top rows around
Mach three and higher angles of attack is due to the noisy depression
area in the data for sideslip level of one-half.
\begin{figure}[ht!]
\centering
\includegraphics[angle=-90,scale=0.35,trim=0 10 10 0]{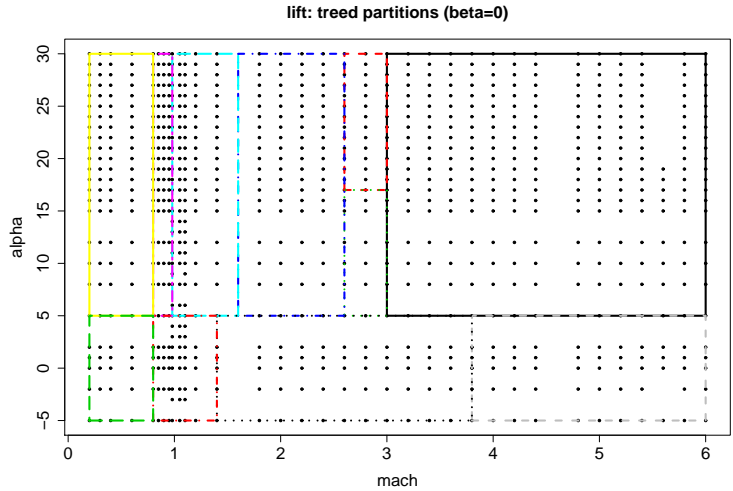} 
\vspace{-0.2cm}
\caption{MAP treed partitions $\hat{\mathcal{T}}$ for the lift response at
sideslip level zero.}
\label{f:lgbb:mapT}
\vspace{-0.2cm}
\end{figure}
Figure \ref{f:lgbb:mapT} shows the MAP treed partitions
$\hat{\mathcal{T}}$ found during MCMC for the slice of sideslip level
zero.  Notice the aggressive partitioning near Mach one due to the
regime shift between subsonic and supersonic speeds. Extra partitioning
at low speeds and large angles of attack address the singularity
outlined in Figure \ref{f:data0}, and near Mach three due to the
numerical instabilities at sideslip level one-half.

\begin{figure}[ht!]
\centering
\includegraphics[angle=-90,scale=0.29,trim=0 60 0 10]{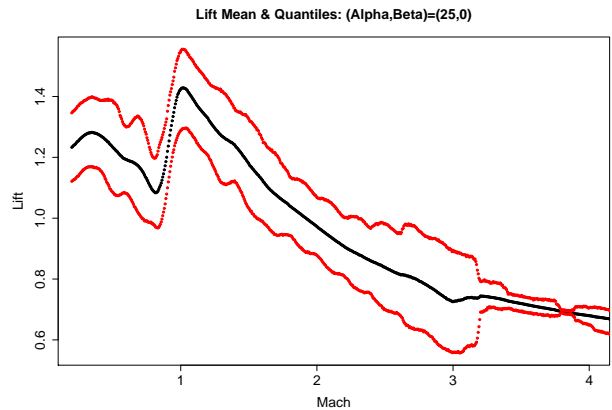} 
\caption{A slice of the mean fit with error bars as a function of Mach with
  alpha fixed to 25 and beta fixed to zero.} 
\label{f:lgbb:mean}
\vspace{-0.3cm}
\end{figure}
Figure~\ref{f:lgbb:mean} shows the mean fit and a 90\% credible
interval for once slice of predicting lift, with Mach on the $x$-axis
and considering only angle of attack equal to 25 and slideslip angle
equal to 0 (i.e., this is one slice from the upper left plot in
Figure~\ref{f:lgbb:surf}; this plot is from fitting the whole dataset,
but we only plot one slice for visibility).  The key item to note is
that the fit is essentially continuous.  The plot is made up only of
points at fitted values, no interpolation or lines have been used. 
\begin{figure}[ht!]
\centering
\vspace{-0.75cm}
\includegraphics[scale=0.6,angle=-90,trim=0 0 74 30,clip=TRUE]{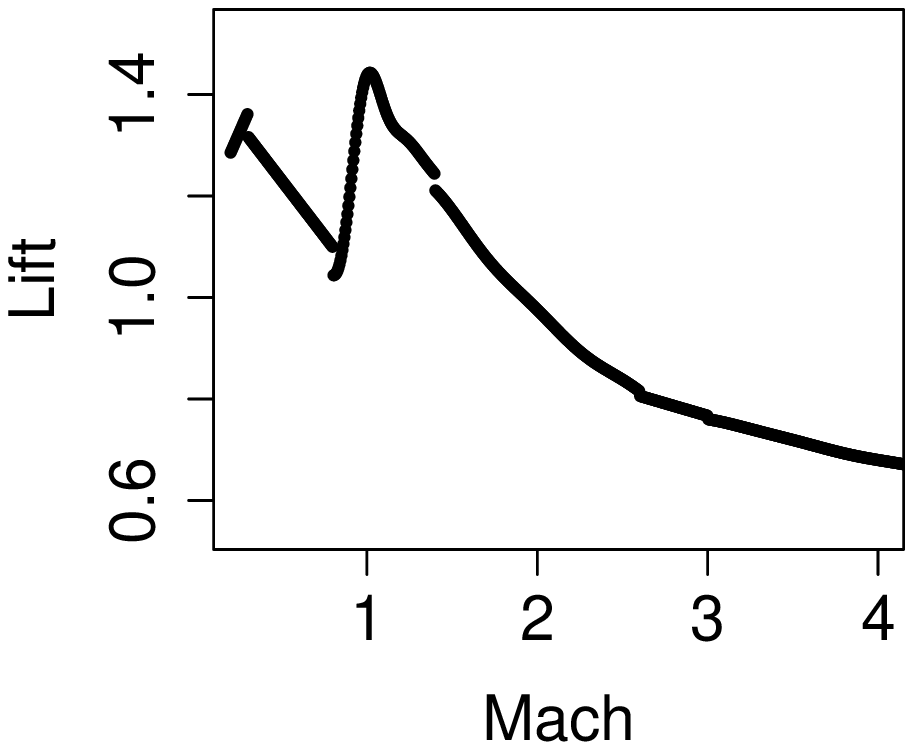}
\includegraphics[scale=0.6,angle=-90,trim=0 58 74 30,clip=TRUE]{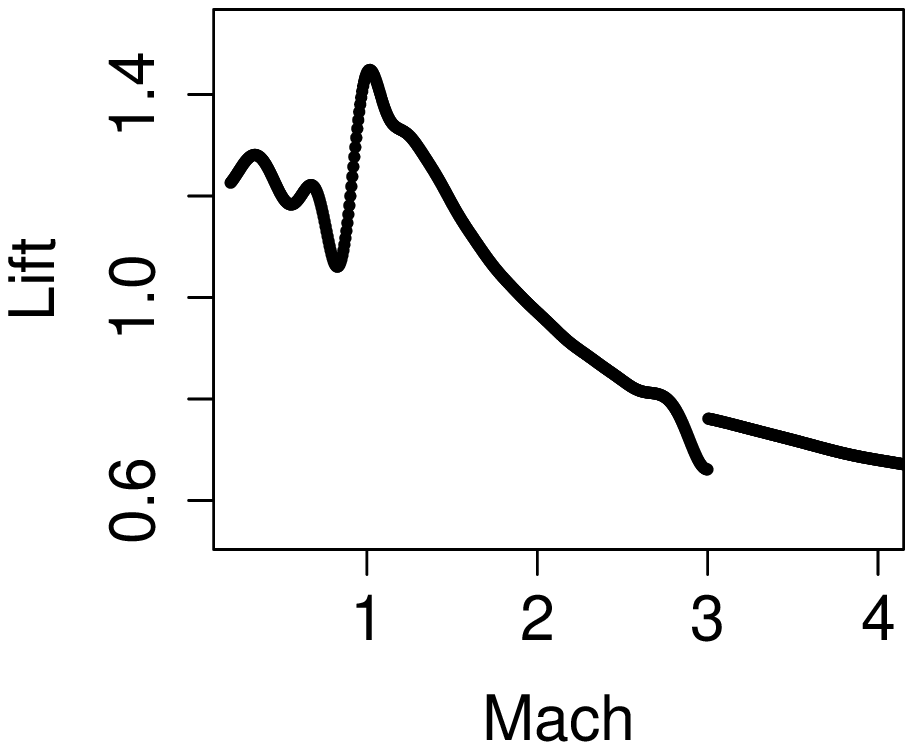}
\includegraphics[scale=0.6,angle=-90,trim=0 58 74 30,clip=TRUE]{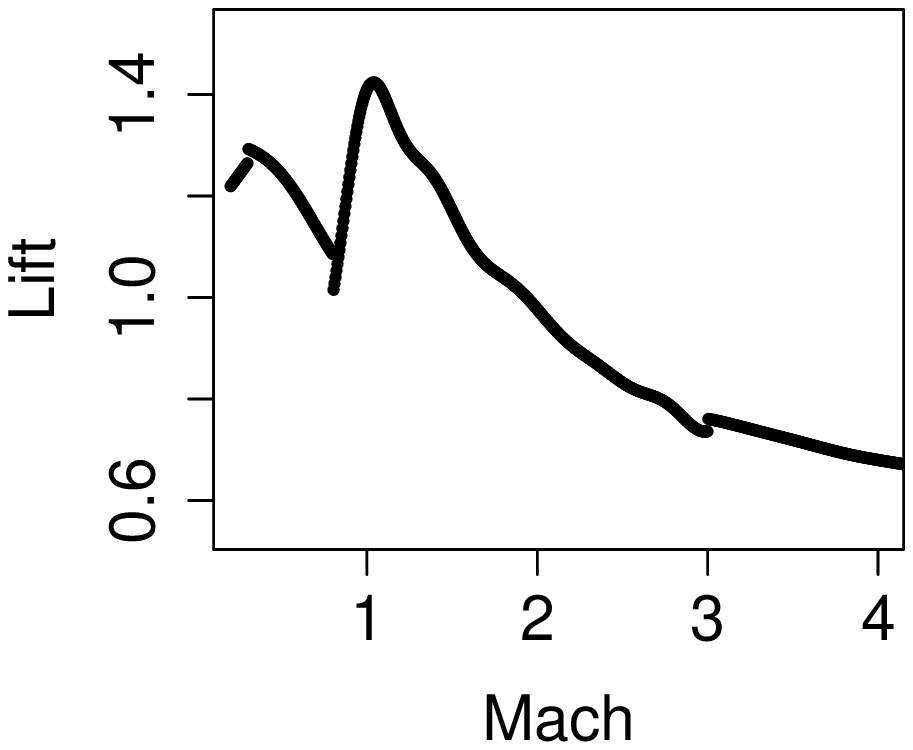}
\includegraphics[scale=0.6,angle=-90,trim=53 0 0 30,clip=TRUE]{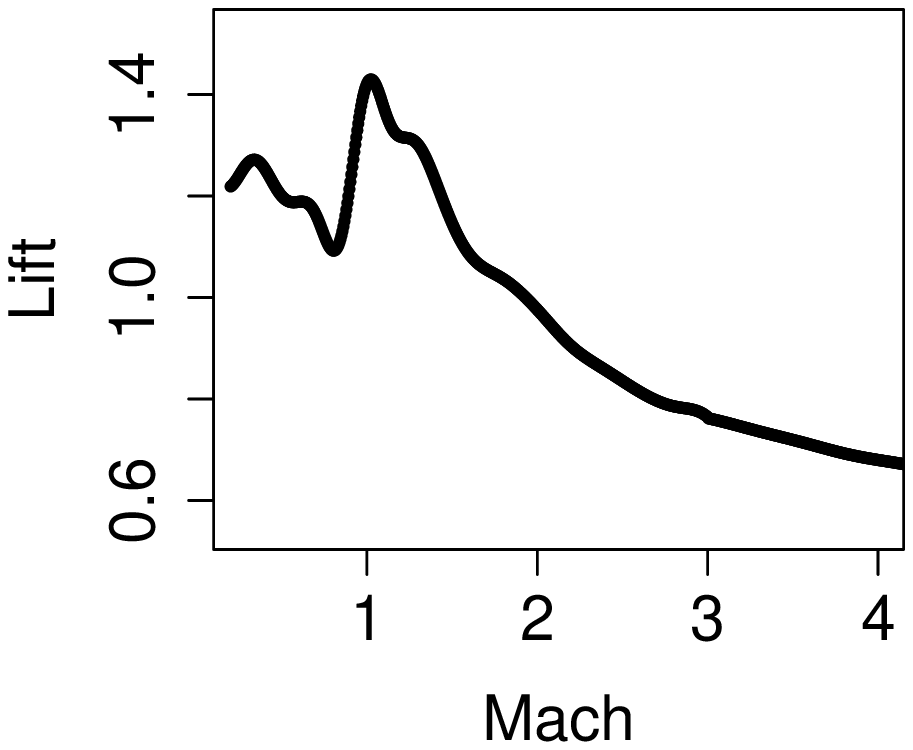}
\includegraphics[scale=0.6,angle=-90,trim=53 58 0 30,clip=TRUE]{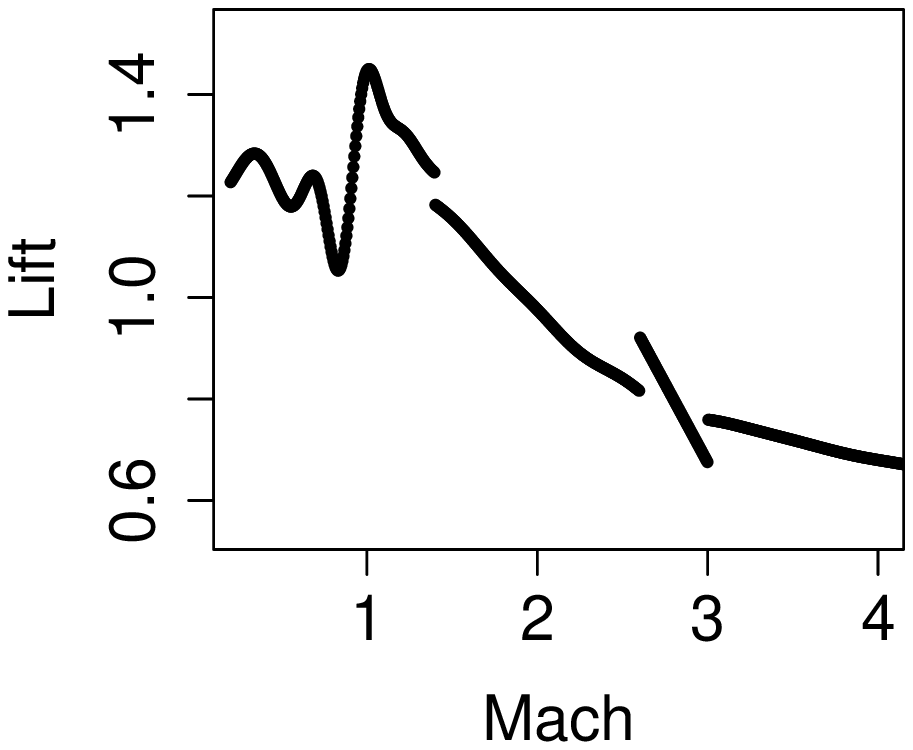}
\includegraphics[scale=0.6,angle=-90,trim=53 58 0 30,clip=TRUE]{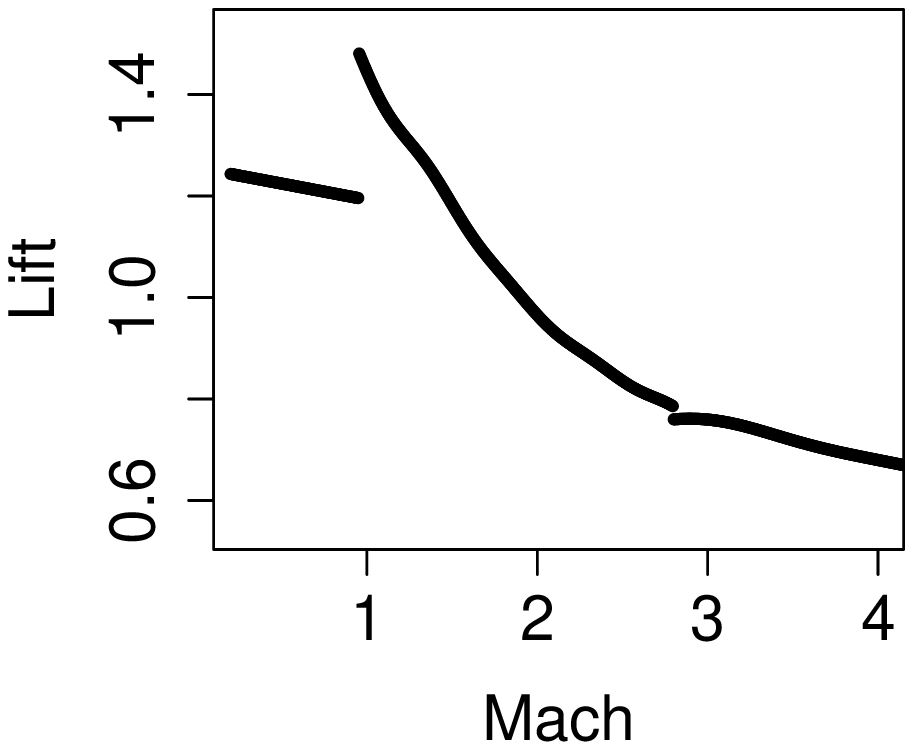}
\vspace{-0.3cm}
\caption{Slices of the posterior predictive mean from individual MCMC
  iterations for the LGBB data with alpha fixed to 25 and beta fixed to zero.
  \label{f:lgbb:samples}}
\end{figure}
In contrast, Figure \ref{f:lgbb:samples} shows examples of the treed
GP fits from individual MCMC iterations, which often have clear
discontinuities from the partitioning structure.  Thus as is typical,
our mean fitted values are quite smooth, because they are an average,
even though the individual components of the mean may not be continuous.

To measure goodness of fit we typically rely on qualitative visual
barometers.  For example, traces are used to asses mixing in the
Markov chain, and posterior predictive slices and projections are
inspected, as described above.  For a more quantitative assessment we
follow the suggestion of \cite{gelfand:1995} and use 10-fold cross
validation.  Posterior predictive quantiles are obtained for the input
locations held-out of each fold, and the proportion of held-out
responses that fall within the 90\% predictive interval is recorded.
For the LGBB data we found a proportion of 0.96 using the treed GP LLM
model.  Thus our model fits well, and if anything, our predictive intervals
are slightly wider than necessary, so we appear to be fully accounting
for uncertainty.

%

\vspace{-0.3cm}
\section{Conclusion}
\label{sec:conclude}
\vspace{-0.2cm}

We developed the treed Gaussian process model for the rocket booster
computer experiment, but it also has a wide range
of uses as a simple and efficient method for nonstationary modeling.
A fully Bayesian treatment of the treed GP model was laid out,
treating the hierarchical parameterization of the correlation function
$K(\cdot,\cdot)$ as a modular component, easily replaced by a
different family of correlations.  The
limiting linear model parameterization of the GP can be both useful
and accessible in terms of Bayesian posterior estimation and
prediction, resulting in a uniquely nonstationary, semiparametric,
tractable, and highly accurate model that contains the Bayesian treed
linear model as a special case.

We believe that a large contribution of the treed GP will be
in the domain of sequential design of computer experiments
\citep{sant:will:notz:2003,glm:04}.  Empirical evidence suggests that
many computer experiments contain much linearity, as we have seen with
large regions of the space for the rocket booster simulator.  The
Bayesian treed GP provides a full posterior predictive
distribution (particularly a nonstationary and thus region-specific
estimate of predictive variance) which can be used towards active
learning in the input domain.  Exploitation of these characteristics
should lead to an efficient framework for the adaptive exploration of
computer experiment parameter spaces.

\renewcommand{\baselinestretch}{1.6}\small\normalsize

\vspace{-0.4cm}
\bibliography{btgpm5}
\bibliographystyle{jasa}

\end{document}